\documentclass[manuscript]{aastex}
\usepackage{emulateapj5}

\shortauthors{von Braun et al.} \shorttitle{Photometry Results for
M10 and M12 }
\begin{document}

\title{Photometry Results for the Globular Clusters M10 and M12: Extinction Maps,
Color-Magnitude Diagrams, and Variable Star Candidates}

\author{Kaspar von Braun, Mario Mateo, Kristin Chiboucas, Alex Athey}
\affil{University of Michigan} \affil{Department of Astronomy}
\affil{500 Church} \affil{Ann Arbor, MI 48109-1090}
\email{kaspar@astro.lsa.umich.edu, mateo@astro.lsa.umich.edu,
kristin@astro.lsa.umich.edu, alex@astro.lsa.umich.edu}

\and{}

\author{Denise Hurley-Keller}
\affil{Case Western Reserve University} \affil{Astronomy
Department} \affil{10900 Euclid Avenue} \affil{Cleveland, OH
44106} \email{denise@smaug.astr.cwru.edu }

%--------------------------------------------------------------------------------
\begin{abstract}

We report on photometry results of the equatorial globular clusters
(GCs) M10 and M12. These two clusters are part of our sample of GCs
which we are probing for the existence of photometrically varying
eclipsing binary stars. During the search for binaries in M10 and M12,
we discovered the signature of differential reddening across the
fields of the clusters. The effect is stronger for M10 than for
M12. Using our previously described dereddening technique, we create
differential extinction maps for the clusters which dramatically
improve the appearance of the color-magnitude diagrams
(CMDs). Comparison of our maps with the dust emissivity maps of
Schlegel, Finkbeiner, \& Davis (SFD) shows good agreement in terms of
spatial extinction features. Several methods of adding an $E_{V-I}$
zero point to our differential maps are presented of which isochrone
fitting proved to be the most successful. Our $E_{V-I}$ values fall
within the range of widely varying literature values. More
specifically, our reddening zero point estimate for M12 agrees well
with the SFD estimate, whereas the one for M10 falls below the SFD
value. Our search for variable stars in the clusters produced a total
of five variables: three in M10 and two in M12. The M10 variables
include a binary system of the W Ursa Majoris (W UMa) type, a
background RR Lyrae star, and an SX Phoenicis pulsator, none of which
is physically associated with M10. M12's variables are two W UMa
binaries, one of which is most likely a member of the cluster. We
present the phased photometry lightcurves for the variable stars,
estimate their distances, and show their locations in the fields and
the CMDs of the GCs.

\end{abstract}
%--------------------------------------------------------------------------------

\keywords{Globular Clusters: individual (M10, M12, NGC 3201);
color-magnitude diagrams; dust, extinction; binaries: eclipsing;
stars: Population II; stars: variable: general}

%--------------------------------------------------------------------------------
\section{Introduction}

Variable stars have historically served as important tools and
``laboratories'' in our understanding of star formation, the
formation of stellar clusters, and the calibration of distance
determination methods. In particular, the study of eclipsing
binary stars (EBs) in a globular clusters (GC) may be used to
obtain a value for the cluster's distance and a constraint
concerning turnoff masses of the GC stars \citep{paczynski96}.

Simply detecting EBs in the fields of GCs and confirming cluster
membership is a straightforward - though data-intensive - task.
These systems expand the relatively meager sample of EBs which are
currently confirmed GC members \citep[see for example][and
references therein]
{mateo1996,mcvean97,rubenstein97,rucinski2000,clement01}. A
statistical evaluation of the number of known member EBs in GCs
can help in the determination of physical quantities such as the
binary frequency in GCs as a parameter in the study of dynamical
evolution of GCs \citep{hut92}.

Member stars of the clusters interact with each other, primarily
toward the core of the GC where the stellar density is much higher
than in the GC halo.  Due to consequent redistribution of stellar
kinetic energies and orbits, the core stars gradually diffuse into the
halo which, as a result, grows in size. At the same time, the cluster
core itself shrinks, and its density will theoretically reach
infinitely high values in a finite period of time \citep[10-20
$t_{relax}$, according to simulations, see][]{bt87}, a phenomenon
known as core collapse.  Binary stars will most likely be located
toward the GC center due to the fact that a) they are more likely to
form in regions of high stellar density, and b) they will sink toward
the core due to their high masses after a few cluster relaxation
times. Since the binding energy of one individual hard
binary\footnote{Hard binaries are systems whose binding energies are
higher than the kinetic energies of the system itself and of the field
star with which it might interact. Consequently, hard binaries tend to
not be disrupted by encounters with other stars, but actually turn
part of their kinetic energy into additional binding energy of the
system.} can be as high as a few to ten percent of the binding energy
of the entire GC, they may act as an energy source (similar to the
nuclear reaction inside the centers of stars) to halt core collapse. A
binary fraction as low as 10\% in the cluster core will suffice to cool
the the central region of the GC by reversing the outward flow of
energy toward the halo \citep{bt87,hut92} and arrest core collapse.

An additional use of simply detecting member-EBs in GC lies in the
calibration of absolute magnitudes of and corresponding distances to W
Ursa Majoris (W UMa) binaries \citep[see also Section
4.3]{rucinski1994,rucinski1995,rucinski2000}.

The simultaneous analysis of photometric and spectroscopic data
for individual EB systems can moreover provide a direct estimate
of the distance to the system \citep{andersen91,paczynski96}, and
thus, if the EB is a GC member, to the GC itself. The main sources
of error in this distance determination are a) the relation
between surface brightness and effective temperature of the binary
and b) the precise determination of the interstellar reddening
along the line of sight to the EB which can vary substantially
across the field of view of the GC. The distance determination
method itself, however, is free of intermediate calibration steps
and can provide direct distances out to tens of kpc. In turn, the
knowledge of the distances to GCs can then be used to calibrate a
variety of other methods, such as the relation between luminosity
and metallicity for RR Lyrae stars. The very same analysis can, in
principle, be used to obtain the Population II masses of the
individual components of the EB system \citep{paczynski96} to
provide a fundamental check of stellar models at low
metallicities.

We are currently undertaking a survey of approximately 10 Galactic
GCs with the aim of identifying photometrically variable EBs
around or below the main-sequence turnoff (MSTO). Our observing
strategy, aimed at detecting binaries in the period range of
approximately 0.1 to 5 days \citep{hut92} consists of repeated
observations of a set of GCs during each night of an observing
run. Multiple runs are helpful in detecting variables with a
period of close to one day (or to a multiple thereof).

M10 (NGC 6254) and M12 (NGC 6218) are two equatorial clusters in
our sample, located at $\alpha_{2000} = 16^{h} 57^{m} 08.9^{s}$
and $\delta_{2000} = -4^{\circ} 05^{'} 58^{''}$ ($l =
15.14^{\circ}$ ; $b = 23.08^{\circ}$), and at $\alpha_{2000} =
16^{h} 47^{m} 14.5^{s}$ and $\delta_{2000} = -1^{\circ} 56^{'}
52^{''}$ ($l = 15.72^{\circ}$ ; $b = 26.31^{\circ}$), respectively
\citep{harris1996}. They are nearby (M10: 4.4 kpc and M12: 4.9
kpc) which makes them attractive targets for monitoring studies.
Both clusters have been probed for the existence of variable stars
in the past \citep[for a summary, see][]{clement01}. Previous
studies searched for luminous variables on the red giant and/or
horizontal branch of the clusters and therefore do not overlap
with the magnitude range covered in this work.

Details on our photometry observations and basic data reductions
are given in Section 2. We discuss how we correct for interstellar
extinction along the line of sight to M10 and M12 in Section 3.
Section 4 contains the description and the results of our search
for variable stars in the clusters including the phased
lightcurves for the variables in the cluster fields and our
estimates concerning distances and cluster membership. The methods
used for Sections 3 and 4 are outlined in detail in two previous
publications, namely \citet[BM01 hereafter]{BM01} and \citet[BM02
hereafter]{BM02} , respectively. Finally, we summarize and
conclude with Section 5.
%--------------------------------------------------------------------------------

%--------------------------------------------------------------------------------
\section{Observations and Data Reductions}

\subsection{Observations}

The observations for M10 and M12 were obtained during multiple
runs between June 1995 and May 1999 at the Michigan-Dartmouth-MIT
(MDM) Observatory's 1.3m and the Las Campanas Observatory (LCO) 1m
Swope Telescope. Tables 1 and 2 list the details. During all
observing runs, we used $2048^2$ pix CCDs with different field
sizes and standard Johnson-Cousins $VI$ filters. The number of
epochs was about evenly divided between $V$ and $I$. During two
runs, we took additional shorter (10s) exposures to complete the
CMDs in the brighter regions (see Tables 1 \& 2). These shorter
observations were not used to look for variable stars because of
infrequent time coverage. All exposures are centered on the
cluster center itself. The reason for the rather large number of
observing runs with relatively few epochs per run (compared to the
other clusters in our sample) is the equatorial location of both
M10 and M12. Consequently, neither cluster is ever located closer
to the zenith than $30^{\circ}$ from MDM as well as from LCO. As a
result, the clusters were observed for only up to $\sim$ 2 hours
each per night when seeing conditions were good, with the rest of
the time dedicated to other clusters in our sample.

\subsection{Data Processing and Reduction}

The details of the IRAF\footnote{IRAF is distributed by the
National Optical Astronomy Observatories, which are operated by
the Association of Universities for Research in Astronomy, Inc,
under cooperative agreement with the NSF.} processing as well as
the basic data reduction of the May 1998 data are described in
BM01, section 2.1. The IRAF processing of the other data sets was
performed in the same manner. All data were reduced using DoPHOT
\citep{schechter93} in fixed-position mode after aligning every
image to a deep-photometry template image consisting of the best
seeing frames for each filter from every run (see BM01). Aperture
corrections and a variable PSF were fitted to the May 1998 data as
second order polynomials in $x$ and $y$. Our astrometric solutions
for both cluster fields, based on approximately 120 US Naval
Observatory (USNO) reference stars \citep{monet96}, respectively,
produced linear fits in $x$ and $y$ with a random mean scatter
(rms) of around 0.2 arcsec, consistent with the USNO precision.
Photometric calibration of the May 1998 data is described in BM01.

All other data were shifted to the coordinates and the photometric
system of the May 1998 600s exposures, including the shorter
exposures. In the process of combining the 10s exposures with the deep
data, we fit for a zero point and color term to match the two sets,
thereby effectively treating the deep data as photometric
standards. Whenever we had two measurements for the same star, we kept
the one with the lower photometric error associated with it (in most
cases, this was the 600s measurement). The division between the two
photometry sets occurs around $V \sim 17-17.5$. We note that we are
sensitive to variable detection to a magnitude as bright as $V \sim
16.5$ since the CCDs we used have different quantum efficiencies and
some of the observations were taken through thin layers of clouds.

Photometric results for every star, weighted by the inverse square of
the DoPHOT photometric error for the respective measurement, were
averaged with the only requirement that a star appear\footnote{ A
detection of a star occurs when a subraster of pixels with a signal
sufficiently high above the sky background is filtered through a
DoPHOT stellar model profile (based on the parameters $x$, $y$,
central intensity, and three shape parameters) and the object under
investigation is classified as a star. } in approximately 20\% or more
of the epochs. This condition is much less stringent than its
corresponding counterpart in the analysis of NGC 3201 (see BM01 and
BM02) where we set this threshold to 75\%. The reason for the
difference lies in the varying field sizes of the CCDs used for M10
and M12 (see Tables 1 \& 2). A larger field (LCO data) will obviously
include more stars than a smaller one, whereas a higher spatial
resolution (MDM data) allows us to get closer to the GC center with
our reductions.

\subsection{Photometry Results and Comparison to Previous Studies}

Our photometry results are presented, in the form of CMDs, in Figures
\ref{M10_rawcmd} (M10) and \ref{M12_rawcmd} (M12). Based on
similarities between the undereddened CMDs of NGC 3201 (presented in
BM01) and M10 it appears that M10 suffers from differential reddening
across the field of view, as previously suggested by, e.g,
\citet{kbk96}. M10's main sequence is broad, especially around the
MSTO where the reddening vector \citep{ccm89,sfd98} is approximately
perpendicular to a hypothetical isochrone fit to this particular
region. Its subgiant branch, giant branch, and horizontal branch (HB)
display the same feature, whereas the fainter region of the main
sequence ($V > 20$), where the reddening vector is almost parallel an
isochrone fit to the region, appears tighter. This behavior clearly
rules out any sort of photometric error as the reason for the
broadness of the CMD features since, in that case, one would observe
well defined MSTO and subgiant branch regions and a main sequence
which would flare up at fainter magnitudes. M12's CMD, does not
display the signature of differential extinction as strongly as M10,
indicating that the effects of differential reddening across its field
are small, despite the fact that the two clusters are separated by
less than $3^{\circ}$ in the sky.

In our photometry we estimate the MSTO of M10 to be around $V-I_{MSTO}
\sim 0.88$ \& $V_{MSTO} \sim 18.5$ (Figure \ref{M10_rawcmd}), and the
one of M12 to be at $V-I_{MSTO} \sim 0.84$ \& $V_{MSTO} \sim 18.35$
(Figure \ref{M12_rawcmd}). Two recent $VI$ studies present themselves
as a means for photometry comparison: the extensive $VI$ GC-CMD
catalog of \citet[RB00 hereafter]{rosenberg00} and the {\it Hubble
Space Telescope (HST)} study of M10, M22, and M55 of \citet[PZ99
hereafter]{pz99}. For M12 we find excellent agreement with RB00 for
our photometry for both $(V-I)_{MSTO}$ and $V_{MSTO}$. For M10 our
MSTO is slightly bluer and brighter than that of RB00 ($(V-I)_{MSTO}
\sim 0.92$ \& $V_{MSTO} \sim 18.6$).  We attribute this to the fact
that the region of M10 observed by RB00 lies to the East of the
cluster center where we find higher differential extinction compared
to the rest of the cluster field (see Section 3.1, and
Figs. \ref{M10_extmap_lowres} and
\ref{M10_grid_lowres_trans}). Plotting our data which fall into the
R00 field results in a location of the MSTO within 0.02 mags of the
R00 MSTO which approaches the limits of the rms of the photometry.

The agreement between our M10 results and the {\it HST} photometry of
PZ99, however, is not as good since their MSTO is at $(V-I)_{MSTO} >
1$. As one can see in Figure \ref{M10_rawcmd}, the range in our data
does not extend to values redder than $(V-I)_{MSTO} \sim 0.95$. Since
a direct star-by-star comparison between our data and PZ99 is not
possible, we cannot give a good explanation for this discrepancy.  One
possible reason, however, might be a zero point drift in the HST
photometry which has been noticed in studies such as \citet{gfm99},
especially since PZ99's photometry zero point is based on HST data.

In addition, we compared our MSTO magnitude to two $BV$ studies of
M10 \citep{hrf89} and M12 \citep{srf89}. The $V_{MSTO}$ for both
M10 and M12 is estimated to be at 18.4, in good agreement with our
results.

%--------------------------------------------------------------------------------

%--------------------------------------------------------------------------------
\section{Dereddening}

\subsection{Differential Extinction}

As mentioned above, the calibrated CMDs in Figures
\ref{M10_rawcmd} and \ref{M12_rawcmd} show the effects of
differential reddening across the respective field of view. In
order to get precise magnitudes and intrinsic colors for potential
GC-member binaries in the clusters, we created differential
extinction maps for M10 and M12 from our data for these clusters.
The method we used is outlined in detail in BM01, section 2.2 and
figures 1-4. We will briefly review the most important steps here
and outline the difference between the method used for NGC 3201
(BM01) and the one used here.

For each cluster, we chose a fiducial region in which little
differential reddening occurred, whose overall extinction is low
compared to the rest of the field, and which contained a
sufficient number of stars to fit a polynomial to the main
sequence for stars with $18.5 < V < 21$ and $0.7 < V-I < 1.4$. The
field of view of each cluster was then divided into subregions of
different sizes such that the number of stars in each of the
subregions was large enough to produce a statistically significant
result. For each of the subregions, an average offset between its
stars and the polynomial fit along the reddening vector with slope
2.411 \citep[table 6, column 4 for Landolt filters]{sfd98} was
calculated after $1 \sigma$-outliers were deleted with an
iteration. This calculated average offset corresponds to the
differential $E_{V-I}$ of each star in the subregion with respect
to the fiducial region, i.e., how much more extinction the stars
in the subregions suffer than the ones in the fiducial region.

The slope of the reddening vector for the analysis of NGC 3201 in BM01
was 1.919, calculated using table 3 in \citet{ccm89}. The central
wavelength of the $I$ filter in \citet{ccm89} is around 9000 \AA,
whereas the filter used during the May 1998 LCO observing run has a
$\lambda_{central} \sim 8100$ \AA, closer to the value given in table
6 of \citet{sfd98} for Landolt filters.  As a result, the conversion
between $E_{B-V}$ to $E_{V-I}$ changes to $E_{V-I} = 1.375 \times
E_{B-V}$, and the slope of the reddening vector changes to
$A_{V}/E_{V-I} = 2.411$. We comment on the implications of this change
in Section 3.2.4.

We present our results for M10 and M12 in the form of differential
extinction maps in Figures \ref{M10_extmap_lowres} (M10) and
\ref{M12_extmap_lowres} (M12). North is up and East is to the left
(see coordinate grid on the figure included for reference). Dark
corresponds to regions of higher extinction relative to the
fiducial region. For reference we show the approximate locations
of the cluster center and core radii \citep[from][]{harris1996} as
well as the fiducial regions of the clusters. The sizes of the
pixels in these maps vary with stellar density from $\sim$ 140
arcsec on the side towards the center of the field to $\sim$ 280
arcsec on the side in the outer regions of the CCD. Compared to
our analysis of NGC 3201 (BM01) the resolution of these extinction
maps is lower by a factor of four (in area) due to the lower
cluster-star density in these pixels.

The average $E_{V-I}$ relative to the fiducial region (field minus
fiducial) is $68 \pm 41$ mmag for M10 and 40 $\pm$ 57 mmag for M12;
the errors represent the standard deviation about the mean (BM01). The
individual subregions' average $E_{V-I}$ values in millimagnitudes
relative to the fiducial region are shown as the top number in each of
the pixels in the grids of Figures \ref{M10_grid_lowres_trans} and
\ref{M12_grid_lowres_trans}. The bottom number in each of the pixels
corresponds to our estimate for the error in the mean for the
corresponding $E_{V-I}$ value (see BM01, section 2.2, item 7 for a
description of our error analysis). The location of the fiducial
region is shown in grey in each of the two Figures. We further show
the division between an outer ``ring" of subregions and the inner part
of the field. The inner subregions in each of the fields typically
contain higher numbers of stars and have lower errors associated with
them. The stellar density is especially low towards the corners of the
fields. The average $E_{V-I}$ for only the inner part of the field
compared to the fiducial region is $49 \pm 28$ mmag for M10 and $9 \pm
15$ mmag for M12.

To illustrate the effects of our internal dereddening method, we
show, in Figures \ref{M10_dered_res3} and \ref{M12_dered_res3},
the CMDs after applying the extinction maps (differential
reddening correction) to the data. The data points are now all
shifted to the CMD-location of the stars in the respective
fiducial region, that is, no absolute reddening zero point has
been applied. The improvement is clearly visible in both cases and
especially obvious for M10 (compare with Figures \ref{M10_rawcmd}
and \ref{M12_rawcmd}). The width of the main sequence of Fig.
\ref{M10_dered_res3} has, by applying the differential reddening
map, decreased to a fraction of its former value in Fig.
\ref{M10_rawcmd}. The subgiant, giant, and horizontal branches
have become much more defined. Even M12's main sequence is
significantly tighter in Figure \ref{M12_dered_res3}, and the
scatter of the stars about its subgiant and horizontal branches is
much lower than in Figure \ref{M12_rawcmd}. In both differentially
dereddened CMDs the small flaring of the data points at the faint
end of the giant branch is most likely due to the low
signal-to-noise of the 10s exposures at these magnitudes.

\subsection{Reddening Zero Point Determination}

At this point, our extinction values for M10 and M12 are
differential, that is, they show $E_{V-I}$ with respect to a
fiducial region in the respective cluster field. This fiducial
region itself suffers some mean interstellar extinction. In order
to a) make our reddening maps a useful tool, and b) determine
intrinsic magnitudes and colors for binary system GC-member
candidates, we need to determine this reddening zero point to add
to our differential $E_{V-I}$ values. As described in BM01, this
cannot be done using our data alone. We therefore use results from
previous studies, as described below. We note that a direct
comparison with literature values should be taken with caution
since usually only a single numerical value for an average
$E_{V-I}$ per cluster is given whereas our result is a map of
extinction across the field of the cluster.

\subsubsection{Using Basic Isochrone Fitting}

The most straightforward method of determining the extinction zero
point is fitting isochrones to our differentially dereddened data (as
performed in BM01) . Thus, the obtained $E_{V-I}$ corresponds to the
reddening zero point to add to the differential $E_{V-I}$-values (see
BM01, section 4.2.1). We performed basic isochrone fitting using a set
of isochrones provided by Don VandenBerg \citep[D. VandenBerg 2000,
private communication, based on evolutionary models by][hereafter
VDB]{vdb00}. Simultaneously fitting age, distance, and extinction
produced the fits shown in Figures \ref{M10_isocmd} (M10) and
\ref{M12_isocmd} (M12). For an estimate of [Fe/H], we used isochrones
with values of [Fe/H] ranging from --1.14 to --2.31, straddling the
cluster [Fe/H] values in \citet{harris1996} of --1.52 for M10 and
--1.48 for M12.

\subsubsubsection{M10}

The isochrone fit for M10, shown in Figure \ref{M10_isocmd}, was
obtained using the following parameters: [Fe/H] = --1.54, age = 16
Gyrs, d = 5.1 kpc ($V_{0} - M_{V} = 13.55$), and the $E_{V-I}$
zero point, to which any differential extinction has to be added,
is 230 mmag. All shapes of the CMD features are well traced out by
the VDB isochrone. The 16 Gyr isochrone produced a slightly better
fit than the 18 Gyr one, but from the appearance of the CMD, it
seems that perhaps a 17 Gyr VDB isochrone, if it were available,
would produce an even better fit than the one shown here. The
value for [Fe/H] is similar to the \citet{harris1996} one of
--1.52, whereas our distance estimate of 5.1 kpc is slightly
higher than the \citet{harris1996} one of 4.4 kpc.

\subsubsubsection{M12}

Since M12's undereddened CMD is very similar to the one of M10,
apart from the much lower differential reddening, it is no
surprise that the parameters for the VDB isochrone fit shown in
Figure \ref{M12_isocmd} are almost identical to the ones used for
M10: [Fe/H] = --1.54, age = 16 Gyrs, d = 4.9 kpc ($V_{0} - M_{V} =
13.46$), and the $E_{V-I}$ zero point is 240 mmag. Again all
features of the CMD, including the subgiant branch and the RGB,
are very well traced out. The only minor deviation between the
loci of the data points and the isochrone fit occurs at $V \sim
21.5$. The above isochrone parameters agree with the values for
M12 in \citet{harris1996}.

\subsubsection{Using SFD Maps}

A second option in the determination of the reddening zero point
is to simply calculate the average offset between our maps and the
SFD dust maps. If the two maps trace out similar dust features,
the offset between the maps should approximately be constant
everywhere in the field. This average offset then corresponds to
the estimate of the reddening zero point based on the comparison
between our extinction maps and the SFD maps (based on dust
emissivity).

We show the pixel-interpolated SFD maps of our fields of view of
M10 and M12 in Figures \ref{M10_sfd_lowres} and
\ref{M12_sfd_lowres}. The orientation and field sizes of Figures
\ref{M10_sfd_lowres} and \ref{M12_sfd_lowres} are identical to the
ones in Figures \ref{M10_extmap_lowres} and
\ref{M12_extmap_lowres}, respectively. As before, darker regions
correspond to regions of higher extinction. We added the location
of the fiducial region for reference in each Figure. For M10, the
average reddening of the SFD map is $E_{V-I}=387 \pm 17$ mmag for
the whole field and $389 \pm 16$ mmag for the inner region (see
Figure \ref{M10_grid_lowres_trans}). The errors represent the
standard deviation about the mean. For M12, these values are
$E_{V-I}=254 \pm 8$ mmag for the whole field and $250 \pm 7$ mmag
for the inner region. Note that these numbers are estimates for
the full extinction along the line of sight, i.e., they contain a
reddening zero point. The SFD estimates agree well with
\citet{harris1996} who quotes $E_{V-I}=382$ mmag for M10 and
$E_{V-I}=258$ mmag for M12.

As a first step in the calculation of the average offset between
the two independent reddening estimates, we subtracted our
extinction maps from the SFD maps of the corresponding regions.
The resulting images are shown in Figures \ref{M10_diffmap_lowres}
(M10) and \ref{M12_diffmap_lowres} (M12). As in BM01, figure 7,
darker regions correspond to areas where our maps indicate more
differential extinction (relative to the fiducial region) than the
corresponding SFD map. We show the locations of the fiducial
regions as well as the border between inner and outer regions (see
above) for reference.

\subsubsubsection{M10}

The difference map for M10 (Figure \ref{M10_diffmap_lowres})
indicates that our extinction map and the SFD map for M10
spatially agree quite well, particularly in the inner region of
the image which basically shows no residual features at all. The
determination of the differential extinction towards the edges of
the map is less precise due to the lower number of cluster stars
in these regions (see Figure \ref{M10_grid_lowres_trans}) and thus
appears more noisy. The average reddening of the whole field of
Figure \ref{M10_diffmap_lowres} (the difference map) is $E_{V-I} =
319 \pm 35$ mmag, whereas for only its inner region, it is
$E_{V-I} = 340 \pm 16$ mmag. It is worth noting that for the inner
region, the rms of this difference map of 16 mmag is smaller than
the rms of our extinction map (28 mmag; see Figure
\ref{M10_extmap_lowres}) and equal to the rms the SFD map for M10
(16 mmag; see Figure \ref{M10_sfd_lowres}), indicating that the
features traced out by both maps are correlated.

The value 340 mmag corresponds to our estimate for the reddening zero
point as determined by using the SFD maps. We were simply unable,
however, to produce an isochrone that resulted in a fit to our data
using this estimate of $E_{V-I}$, even when widely varying the values
for distance, age, and [Fe/H] (see Section 3.2.3 for discussion). We
therefore adopt our estimate of 230 mmag, as determined by the VDB
isochrones, as the reddening zero point for M10 (see previous
Subsection).

\subsubsubsection{M12}

Similar to M10, the two maps for M12 (Figures \ref{M12_extmap_lowres}
and \ref{M12_sfd_lowres}) agree well, resulting in a difference image
(Figure \ref{M12_diffmap_lowres}) which is featureless except in the
regions towards the corner of the image. The average reddening of the
entire difference map (Figure \ref{M12_diffmap_lowres}) is $E_{V-I} =
214 \pm 53$ mmag, whereas for only its inner region, it is $E_{V-I} =
241 \pm 11$ mmag. The value 11 mmag as the rms is comparable to the
rms of the SFD map (7 mmag; see Figure \ref{M12_sfd_lowres}) and lower
than the rms for our extinction map (15 mmag; see Figure
\ref{M12_extmap_lowres}), again indicating positive correlation
between the features present in both maps. We calculate 241 mmag to be
our estimate for the reddening zero point as determined by using the
SFD maps. This value agrees very well with our results from the VDB
isochrone fitting method (240 mmag; see above).

\subsubsection{Comments on the Adopted Reddening Zero Point}

The two independent estimates we obtain for an average $E_{V-I}$
(zero point plus differential) for each cluster are 389 mmag (by
adding the SFD zero point to the average level of our extinction
map) and 279 mmag (by using the VDB isochrone) for M10, and 250
mmag (using SFD zero point) and 249 mmag (using VDB isochrones)
for M12. The reddening estimates in the literature for the two
clusters vary more for M10 than for M12. Our results, along with
some recent $E_{V-I}$ values for M10, are presented in Table 3,
and for M12 in Table 4. It should be noted that all the literature
values in Table 3 \& 4 were obtained by converting their cited
$E_{B-V}$ values to $E_{V-I}$ by using $E_{B-V} = 0.7273 \times
E_{V-I}$ and an $R_{V} = 3.1$ reddening law \citep{sfd98}.

As mentioned above we were not able to fit an isochrone to our
data using the SFD reddening zero point for M10, even when
lowering the metallicity to --2.31 (limit of the isochrones) and
changing the values for the age of the isochrone or the distance
of the cluster. In order to avoid the possibility of multiple
factors conspiring against us (such as incorrect aperture
corrections and/or calibration zero points), we compared our
DoPHOT photometry results for 10 isolated stars (no close
neighbors) with IRAF's qphot magnitudes for the same stars. Within
rms photometry errors, the results were identical, such that we
found an average offset between $I_{DoPHOT} - I_{qphot} \sim
0.034$ and $V_{DoPHOT} - V_{qphot} \sim 0.005$.

We will therefore adopt 230 mmag as the reddening zero point for
M10. Compared to literature values, we are thus below the SFD,
\citet{hrf89}, and \citet{harris1996} estimates, and slightly
above the values in \citet{bh82} and \citet{amp90}. For M12, our
reddening estimate agrees very well with the SFD and
\citet{harris1996} values, but is slighter higher than the
\citet{bh82} and slightly lower than the \citet{srf89} estimates.

The similarity in our $E_{V-I}$ values for the clusters may be
somewhat unexpected due to the strong presence of differential
reddening across the field of M10 which is not observed in M12. On
the other hand, the two observed (undereddened) CMDs of the
clusters are extremely similar to each other \citep[confirmed also
in][]{hrf89,srf89,rosenberg00}. Given the small difference between
the values of [Fe/H] for M10 and M12, it therefore appears quite
sensible that the estimates for age, distance, and reddening
should be very similar for the two GCs.

\subsubsection{Comments on the BM01 Reddening Zero Point for NGC 3201}

As we mentioned in Section 3.1, we changed the value for the slope
of the reddening vector from $A_{V}/E_{V-I} = 1.919$ in BM01,
based on \citet{ccm89}, to 2.411 for this work, based on
\citet{sfd98}. The central wavelength for the $I$ filter used to
establish the \citet{ccm89} conversions is 9000 \AA, whereas the
filter used for the SFD conversions peaks closer to 8100 \AA. Any
instrumental $I$ magnitude will suffer higher extinction with
shorter $I$ filter central wavelengths. Since reddening is not
calculated until after the transformation to standard magnitudes
is performed, the slope of the reddening vector has to be chosen
according to the characteristics of the $I$ filter. The $I$ filter
used during the LCO 1998 observing run has a central wavelength
around 8200 \AA, closer to the one used in SFD.

A recalculation of the reddening maps based on this ``new" slope,
however, produced results which were identical to the previously
calculated ones. The values in the individual pixels of the
extinction map grid (such as in figure \ref{M10_grid_lowres_trans}
and \ref{M12_grid_lowres_trans}) changed by only a few mmag from
the values calculated in BM01. Similarly, refitted isochrones
produced an identical reddening zero point. We therefore conclude
that our reddening results for NGC 3201 (BM01) do not change as a
result of this different slope of the reddening vector .

As another consequence of adopting a different central wavelength
for the $I$ filter, however, the conversion between $E_{B-V}$ and
$E_{V-I}$ changes from $E_{B-V} = 1.546 \times E_{V-I}$ for
$\lambda_{central} \sim 9000$ \AA{} to $E_{B-V} = 1.375 \times
E_{V-I}$ for $\lambda_{central} \sim 8100$ \AA. Thus, literature
$E_{B-V}$ values will correspond to $E_{V-I}$ values which are
different from the ones quoted in BM01. We show these updated
literature values in Table 5. Qualitatively, the conclusions we
reached in BM01 remain the same. Quantitatively, the agreement
between our estimate for $E_{V-I}$ and the corresponding
literature values is better than we stated in BM01. Our value
falls less than 1 $\sigma$ below the \citet{c84} and
\citet{harris1996} values (we quoted 1.5 $\sigma$ in BM01).

%--------------------------------------------------------------------------------

%--------------------------------------------------------------------------------
\section{The Search For Variable Stars}

The starting point for our analysis of the photometry data with
respect to finding binaries and determining their periods were two
databases. One contained the data on approximately 70
observational epochs per filter for M10 (600s exposure time each)
of 20000 stars per image, the other the equivalent data for M12
consisting of about 100 600s epochs per filter with $\sim$ 14000
stars per image. Our data are sensitive to variables as bright as
$V \sim 16.5$, slightly brighter than the previously mentioned
division between 600s and 10s data in the CMDs, since observing
conditions and CCD quantum efficiencies may vary between epochs or
observing runs. Finally, we note that the variability detection
and the period determination of variable stars are entirely
independent of the differential dereddening described in Section
3.

\subsection{Variability Detection}

The criteria we set in order to extract variable star candidates
from the list of stars in our database were described in detail in
BM02, section 3.1. We will only briefly reiterate them here and
mention any differences between BM02 and this work:

\begin{enumerate}
\item $\chi^{2}$ per degree of freedom, calculated based on the
assumption that every star is a non-variable, has to be greater
than 3.0. We furthermore set a $\sigma >$ 0.05 mag threshold for a
star to be taken into consideration as a variable candidate, where
$\sigma$ represents the variability in the lightcurve.
\item The star under investigation must appear in at least 40\% of the
epochs. The reason why this number is considerably lower than it
was set in the analysis of NGC 3201 (BM02) is the same as in
Section 2.2: the MDM data, which comprise a large fraction of the
epochs (see Tables 1 \& 2), were taken with CCDs with smaller
field sizes than the LCO data.
\item The detected variability should not be due to only a few outliers
(3-5\% of the datapoints) causing the high $\chi^{2}$. Care was
taken to avoid deleting possible detached eclipsing systems at
this stage as we outlined in BM02, section 3.1.
\item The star under investigation should display a brightness variation
in both filters, and the variability signal should be correlated
in both filters \citep[this algorithm is very similar to the one
described in][]{WS93,S96}.
\end{enumerate}

Any measurement of stellar magnitude was weighted by the square of
the inverse photometric error associated with it.

\subsection{Period Determination}

The final decision as to whether a star is a true variable
candidate (and if so, what kind of variable star it is) was based
on the inspection of the data phased by the correct period (the
photometry lightcurve). A variety of algorithms to determine
periods are used in astronomy, many of which are based on similar
principles. Generally, a test statistic is defined as a function
of the observations (random variables), and of a trial period
(parameter of the statistic) by which the real-time observations
are folded. For each trial period and a given set of observations
the test statistic returns a single number. One may therefore plot
the value of the test statistic against trial period (or its
corresponding frequency) in what is called a periodogram
\citep{SC89}. Periodic oscillations in the observations will show
up as features (resembling spectral lines) in the periodogram.

The initial estimates of the periods of all our variable star
candidates which survived the steps outlined in the previous
Subsection were determined by two independent algorithms:

\begin{itemize}
\item the minimum-string-length method, based on a technique by
\citet{LK65} and described in \citet{S96}. The test statistic in
this method is the length of a hypothetical piece of string
connecting consecutive data points in a magnitude vs phase plot.
For an incorrect period estimate, the data points will most likely
be scattered, resulting in a long string length. When the trial
period is very close to the true period, however, the string
length will be short since any two data points consecutive in
phase will be very close to each other in magnitude.
\item the Analysis of Variance (AoV) method, described in detail in
\citet{SC89}\footnote{This method is quite similar to the Phase
Dispersion Minimization method}. The basic code for this algorithm was
supplied to us by Andrzej Udalski (private communication, 1998). For
this method, the entire dataset is folded by a test period and then
divided into bins in phase. The test statistic is the ratio of two
variances. The numerator is the variance of the averages of the bins
about the mean of the entire dataset, which obviously is high if a
star is a variable. The denominator is the sum of variances within the
individual bins, which is small if the test period is the correct
period. For a pure noise signal, this ratio (the AoV statistic) is
1. For a periodic signal its value is small for incorrect trial
periods and large for the correct trial period.
\end{itemize}

A final tweaking of the precision in the period for a candidate
was then done by hand whenever necessary, based on the appearance
of the lightcurve.

As in BM02, we note that we did not systematically address the issue
of completeness in our analysis. The choice of parameters for the
steps outlined above was made mainly based on hindsight (e.g., all of
our candidates' $\chi^{2}$ values are well in excess of
100\footnote{Reasons for high $\chi^{2}$ values of non-variable stars
are very low signal and crowding.}) or trial and error (e.g., the
analysis of a set of phased lightcurves before and after a reduction
criterion was applied). Any assessment of incompleteness due to other
effects, such as period, phase, or duty cycle, is difficult to make as
it sensitively depends on the windowing function created by the times
of the observational epochs. If the period of a variable star is less
than about an hour or higher than 5 days, we will most likely not
detect the system since our exposure times are 600s long and observing
runs generally last between one and two weeks. If the phase is such
that the variable will only undergo changes in its brightness during
the daytime, we will obviously also not find it. Very low duty cycles
make it hard to detect the variable since it spends so little time in
an eclipse and otherwise appears as a non-variable star. The lower the
duty cycle, the more likely it is that a combination between the
windowing function and the phasing of the binary would cause us to not
detect it. However, the timing of our observing should enable us to
find all variables with a period of up to around a day with fairly
high duty cycles (30\% or higher) since we perform repeated
observations of the same cluster over the course of hours throughout
many nights (especially for NGC 3201, see BM02).

\subsection{Results}

\subsubsection{Locations of Variable Stars in Field and in CMD}

Table 6 gives the basic information about the variable stars we
detected in the fields of M10 and M12. $V_{bright}$ and
$I_{bright}$ are the $V$ and $I$ magnitudes at maximum light.
Figures \ref{M10_field} and \ref{M12_field} show the locations of
the respective variable stars in the fields of M10 and M12.
Figures \ref{M10_isocmd} and \ref{M12_isocmd}
indicate where the variables fall within the CMDs of the two
clusters. The data shown in Figures \ref{M10_isocmd} and
\ref{M12_isocmd} are differentially dereddened to the
fiducial region within the respective cluster, as described above.
No reddening zero point is applied. The magnitudes and colors of
the variables correspond to the values at maximum brightness.

\subsubsection{Phased Photometry Lightcurves}

The phased lightcurves for the variable stars are presented in Figures
\ref{M10_v1} through \ref{M12_v2}. In particular, Figure \ref{M10_v1}
shows the lightcurve of M10-V1 (period = 0.263366 days) which is a W
Ursa Majoris (W UMa) type contact binary system. M10-V2 is a
background RR Lyrae (ab type) variable with a period of 0.61438 days;
its lightcurve is displayed in Figure \ref{M10_v2}. M10-V3 appears to
be a field $\delta$ Scuti or an SX Phoenicis-type (SX Phe) variable
with a period of 0.0637 days.  These stars are rapidly pulsating
Population I ($\delta$ Scuti) or Population II (SX Phe) stars which
are commonly found towards the faint end of the instability strip in
the CMD \citep{rll00}.  M10-V3's lightcurve is shown in Figure
\ref{M10_v3}. M12-V1, another W UMa binary, seems to be the only
variable in our data set which is physically associated with its
respective parent GC (based on its CMD location).  Its period is
0.243183 days, and its lightcurve is displayed in Figure
\ref{M12_v1}. M12-V2, another W UMa binary, but not associated with
M12, has a period of 0.25212 days; its lightcurve is in Figure
\ref{M12_v2}. In all of these Figures, $V$ data are in the bottom
panel, $I$ data in the top panel. The error bars represent the DoPHOT
photometric errors associated with that particular measurement of the
star's magnitude. No reddening correction was applied to the
lightcurves.

\subsubsection{Variable Stars: Distances and Cluster Membership}

\subsubsubsection{W UMa Binaries (in M10 and M12)}

As indicated in Section 4.3.2, all the binary systems we detect in the
fields of M10 and M12 are of the W Ursa Majoris type. W UMa binaries
are systems in which the two components are in physical contact with
their Roche equipotential lobes and at their inner Lagrangian Point
\citep[see, e.g.,][for more detailed description of W UMa
systems]{rucinski85a, rucinski85b,mateo93}. The fact that the total
brightness of the system is constantly changing by up to $\sim$0.75
magnitudes makes them easy to detect. Due to the good thermal contact
between the two components, the two stars have similar surface
temperatures. The absolute magnitude of the system can therefore be
defined as:

\begin{equation}
M_{V}=-4.43\log P+3.63 (V-I)_{0}-0.31,
\end{equation}

where $M_{V}$ is the absolute $V$ magnitude, $P$ is the period in
days, and the $(V-I)_{0}$ color is reddening-free. The empirically
determined standard deviation in this relation is $\sigma \sim
0.29$, corresponding to about 13\% in distance. See
\citet{rucinski2000} and BM02 for a more detailed derivation of
the above equation.

Table 7 shows the absolute magnitudes and distance moduli for the
W UMa binaries in the fields of M10 and M12 based on the above
relation. The distances to the clusters were calculated in Section
3.2.2 to be 5.1 kpc (M10) and 4.9 kpc (M12). The corresponding
true distance moduli (apparent distance moduli corrected for
extinction) are $V_{0}-M_{V}=13.55$ and 13.46, respectively. We
note that the absolute magnitudes and distance moduli in Table 7
were calculated under the assumption that the W UMa system under
investigation is suffering the full extinction between us and the
cluster. Since the Rucinski magnitudes for some of the variables
indicate that they are foreground stars, this assumption might be
incorrect in some cases. That is, for some of the non-members, the
color might not be the correct, reddening-free value.

A first indication of whether an EB system in the field of a globular
cluster is associated with that cluster is, of course, its location
with respect to the cluster center (we note that we cannot study the
very center of the cluster due to crowding). The tidal radii of M10
and M12 are 21.48 and 17.6 arcmin, respectively
\citep{harris1996}. Thus, all the binary systems we find are well
within the tidal radii. A more powerful membership criterion is a
binary's location in the CMD. Based on the CMDs of M10 and M12 (see
Figures \ref{M10_isocmd} \& \ref{M12_isocmd}), M10-V1 and M12-V2 are
not associated with their respective parent cluster. M12-V1, however,
appears to be a cluster member from its location in M12's CMD. The
estimates of distance moduli to the various W UMa binaries in Table 7
confirm what the GC-CMDs suggest: M10-V1 and M12-V2 are foreground
stars, but M12-V1 is most likely a member of M12 (it falls well within
one $\sigma_{Rucinski}$ of the calculated distance to the GC).

\subsubsubsection{Pulsating Variables (in M10)}

The two pulsating variable stars we detected in M10 are M10-V2, an
ab-type RR Lyrae star, and M10-V3, either a $\delta$ Scuti or an
SX Phe star. From their locations in M10's CMD, it is apparent
that neither of them is associated with the cluster.

Since the ``expected" location of any GC-member RR Lyrae is on the
horizontal branch, M10-V2 is clearly a background star. If one
assumes that $M_{V} \sim 0.7$ for RR Lyraes, M10-V2's distance is
around 40 kpc, putting it into the Milky Way halo beyond the
bulge.

As mentioned above, SX Phe stars represent the Population II
counterparts to the Pop I $\delta$ Scuti variables \citep{rl00}.
GC-member SX Phe stars are typically found in the blue straggler
region of the CMD. Since M10-V3 is not inside this region, we
conclude that it is not a member of the GC. Instead, the
photometry results for M10-V3 ($\Delta V \sim 0.32$, P $\sim$
0.064 days, $(V-I)_{0} \sim 0.29$\footnote{From comparison with
the CMD of M10 by \citet{hrf89}, this approximately corresponds to
$(B-V)_{0} \sim 0.3$, i.e., a spectral type of about F0.}) quite
exactly correspond to the modes in the distributions of the
catalogue of field $\delta$ Scuti and SX Phe stars of
\citet{rll00}; see their figures 1, 3, \& 4. Since the number of
known field SX Phe stars is very low \citep{rll00, rl00, jkl01} we
cannot say for sure whether M10-V3 is a field $\delta$ Scuti or SX
Phe variable \citet[see for example figure 5 in][]{jkl01}. Since
M10-V3 is a background variable, however, and thus located in the
halo, we will assume that it is an SX Phe variable. Using the
appropriate P-L-[Fe/H] relations in \citet{nnl94} and
\citet{nmb95} with an assumed [Fe/H] $\sim$ --1.5, we obtain a
distance to M10-V3 of approximately 16 kpc.

%--------------------------------------------------------------------------------

%--------------------------------------------------------------------------------
\section{Summary and Concluding Remarks}

The GCs M10 and M12 were monitored for the existence of
photometrically variable stars in the magnitude range of
approximately $16.5<V<20$ and for periods between roughly 0.05 and
5 days. Observations were conducted over a span of about four
years with only relatively few epochs per observing run due to the
equatorial locations of the clusters.

\begin{figure}
\epsscale{0.5} \plotone{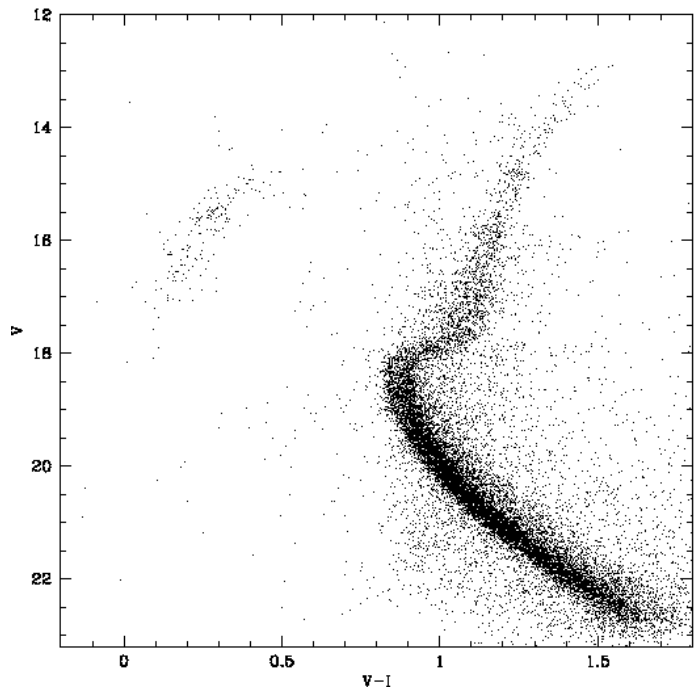}
\caption{\label{M10_rawcmd}CMD of M10 before applying any
extinction correction. The typical symptoms of differential
extinction across the field are present: main sequence appears
relatively tight where it is approximately parallel to the
reddening vector (at $V$ fainter than $\sim20$), whereas regions
more or less perpendicular to the reddening vector, such as the
MSTO, the subgiant and giant branches as well as the blue HB,
appear broad. As part of the process of combining 600s and 10s
exposure data, we set the division at around $V \sim 17-17.5$
where saturation of the deep data sets in. }
\end{figure}

\begin{figure}
\epsscale{0.5}
\plotone{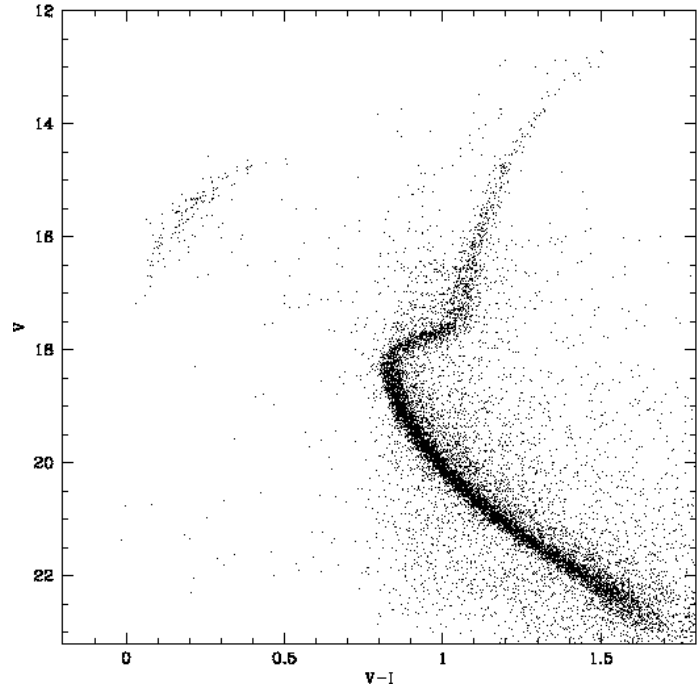}
\caption{\label{M12_rawcmd}CMD of M12 before applying any
extinction correction. The signatures of differential reddening
are much less visible than in M10 (see Fig. \ref{M10_rawcmd}). The
division between the deep (600s) and shallow (10s) photometry data
is located around $V \sim 17-17.5$.}
\end{figure}

\begin{figure}
\epsscale{0.5}
\plotone{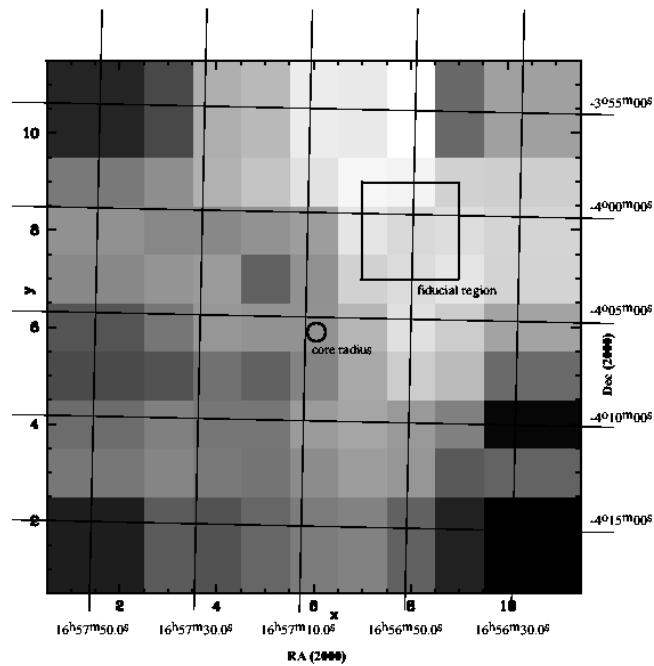}
\caption{\label{M10_extmap_lowres}Extinction Map for M10. North is
up, east is to the left. The size of the entire field is 25.9
arcmin on the side. The darker the color of a subregion, the more
extinction is occurring in it (see Figure
\ref{M10_grid_lowres_trans}) relative to the fiducial region whose
position is shown on the map along with the core radius around the
location of the center of the cluster. The coordinate axes $x$ and
$y$ correspond to the ones in Figure \ref{M10_grid_lowres_trans}.
The average reddening with respect to the fiducial region is
$E_{V-I} = 68 \pm 41$ mmag for the entire field and $E_{V-I} = 49
\pm 28$ mmag for the inner part of the field (see Figure
\ref{M10_grid_lowres_trans}). The size of one pixel in this map
($\Delta x = 1$ or $\Delta y = 1$) is $\sim 140$ arcsec.}
\end{figure}

\begin{figure}
\epsscale{0.5}
\plotone{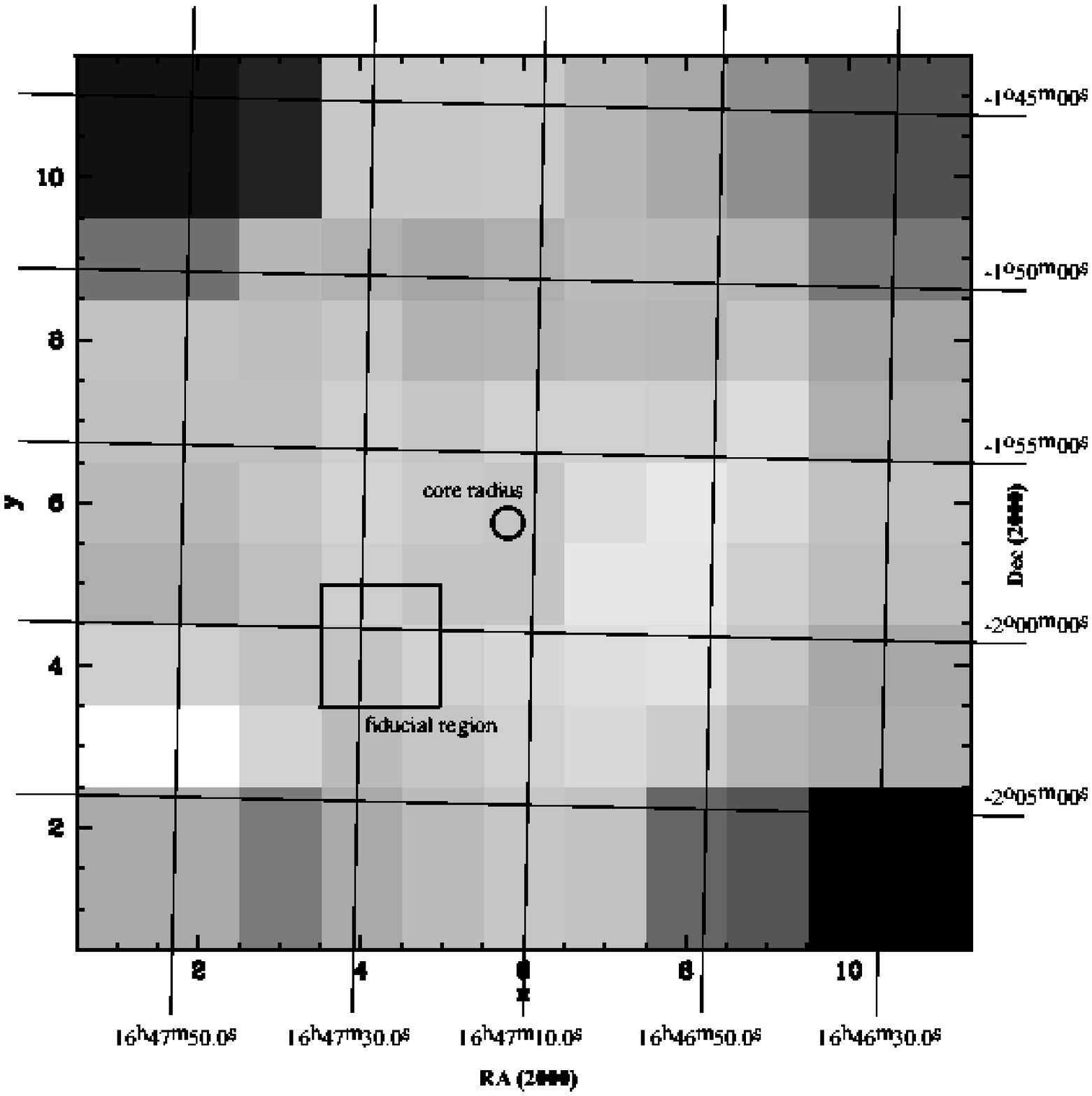}
\caption{\label{M12_extmap_lowres}Extinction Map for M12. North is
up, east is to the left. The size of the entire field is 25.9
arcmin on the side. The darker the color of a subregion, the more
extinction is occurring in it (see Figure
\ref{M12_grid_lowres_trans}) relative to the fiducial region whose
position is shown on the map along with the core radius around the
location of the center of the cluster. The coordinate axes $x$ and
$y$ correspond to the ones in Figure \ref{M12_grid_lowres_trans}.
The average reddening with respect to the fiducial region is
$E_{V-I} = 40 \pm 58$ mmag for the entire field and $E_{V-I} = 9
\pm 15$ mmag for the inner part of the field where the errors are
considerably smaller (see Figure \ref{M12_grid_lowres_trans}) due
to the higher stellar density. The size of one pixel in this map
($\Delta x = 1$ or $\Delta y = 1$) is $\sim 140$ arcsec.}
\end{figure}

During the search for variables, we noticed strong differential
reddening effects across the field of M10 (Figure
\ref{M10_rawcmd}) and a similar, but weaker signature across M12
(Figure \ref{M12_rawcmd}). We correct for this differential
reddening by creating extinction maps (Figures
\ref{M10_extmap_lowres} and \ref{M12_extmap_lowres}). The features
visible in these maps are similar to the ones detected by SFD in
their dust emissivity maps of the same regions (Figures
\ref{M10_sfd_lowres} and \ref{M12_sfd_lowres}). Applications of
our differential extinction maps to our data significantly
improves the appearance of the CMDs of the GCs (Figures
\ref{M10_dered_res3} and \ref{M12_dered_res3}), especially in the
case of M10.

\begin{figure}
\epsscale{0.5}
\plotone{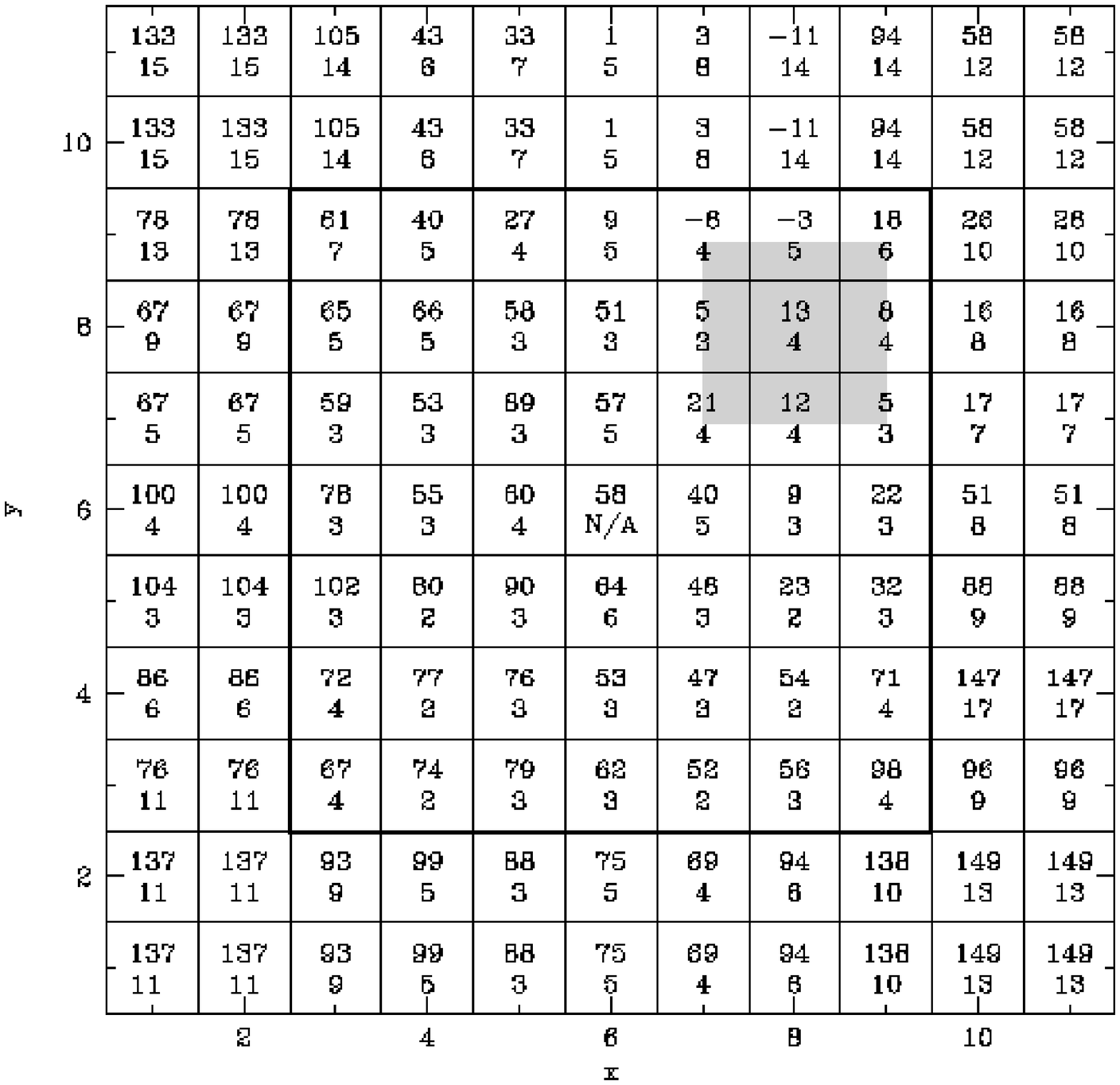}
\caption{\label{M10_grid_lowres_trans}M10 -- Differential
$E_{V-I}$ values (top number) with associated errors (bottom
number) relative to the average $E_{V-I}$ of the fiducial region
(shown in grey for reference). Each pixel is about 140 arcsec on
the side. The size and orientation of the field is the same as in
Figure \ref{M10_extmap_lowres}. The pixel in the map with an error
value listed as N/A did not contain enough stars for a fit, due to
saturation and crowding in the cluster center. Its $E_{V-I}$ was
obtained by interpolation from neighboring pixels. The average
reddening with respect to the fiducial region is $E_{V-I} = 68 \pm
41$ mmag for the entire field and $E_{V-I} = 49 \pm 28$ mmag for
the inner part, shown in the image. The subregions in this inner
part carry higher significance due to the higher number of cluster
stars they contain. To obtain a map of absolute $E_{V-I}$ values,
a reddening zero point has to be added to the pixel values. We
discuss this zero point in Section 3.2. Finally, in order to get
absolute values for $E_{B-V}$, one needs to multiply the absolute
$E_{V-I}$ value (i.e., including zero point) by 0.7273
\citep{sfd98}. }
\end{figure}

\begin{figure}
\epsscale{0.5}
\plotone{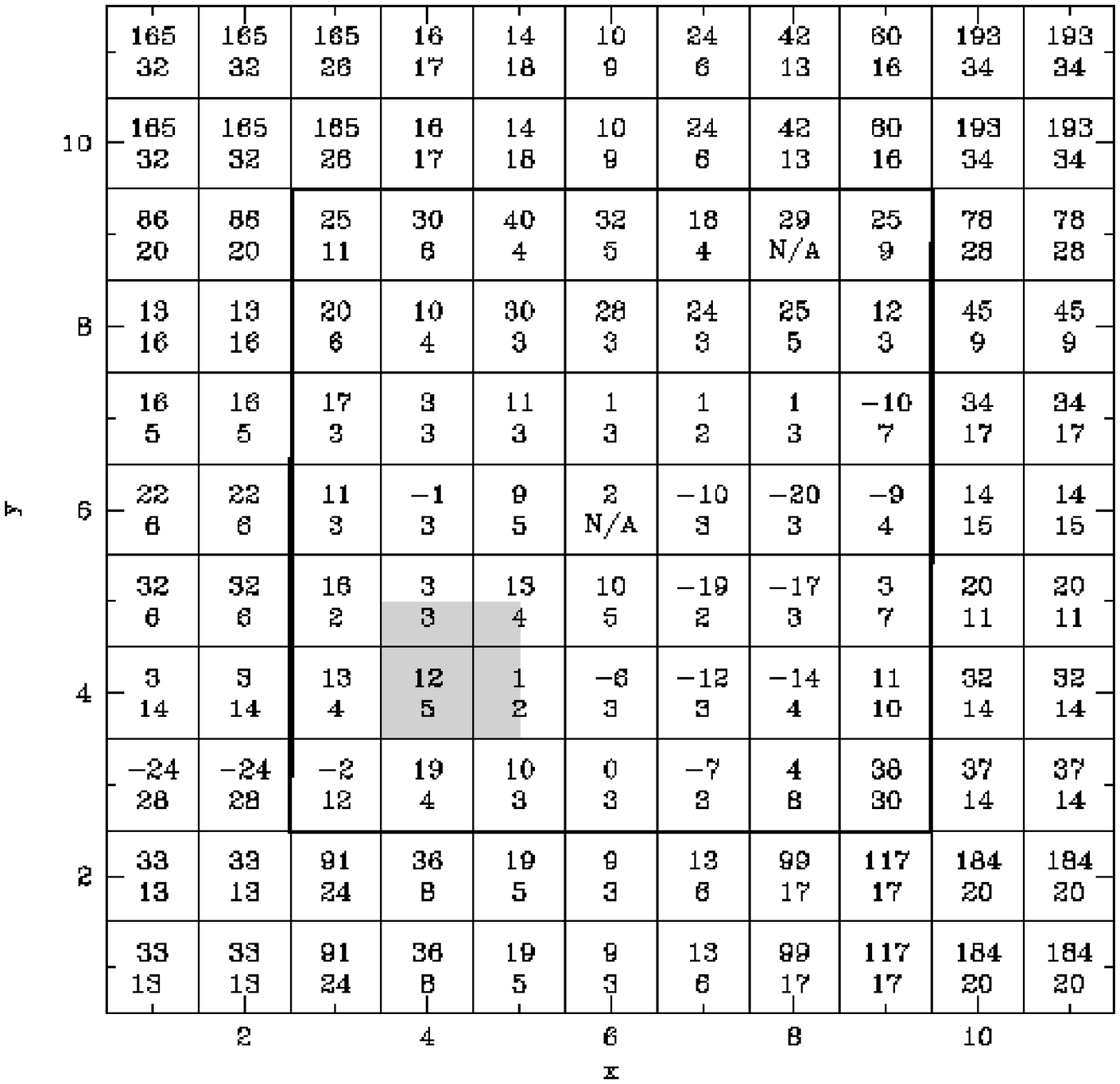}
\caption{\label{M12_grid_lowres_trans}M12 -- Differential
$E_{V-I}$ values (top number) with associated errors (bottom
number) relative to the average $E_{V-I}$ of the fiducial region
(shown in grey for reference). Each pixel is about 140 arcsec on
the side. The size and orientation of the field is the same as in
Figure \ref{M12_extmap_lowres}. The pixels in the map with error
values listed as N/A did not contain enough stars for a fit. Their
$E_{V-I}$ were obtained by interpolation from neighboring pixels.
The average reddening with respect to the fiducial region is
$E_{V-I} = 40 \pm 58$ mmag for the entire field and $E_{V-I} = 9
\pm 15$ mmag for the inner part (shown) of the field where the
errors are smaller. To obtain a map of absolute $E_{V-I}$ values,
a reddening zero point has to be added to the pixel values. We
discuss this zero point in Section 3.2. Finally, in order to get
absolute values for $E_{B-V}$, one needs to multiply the absolute
$E_{V-I}$ value (i.e., including zero point) by 0.7273
\citep{sfd98}. }
\end{figure}

\begin{figure}
\epsscale{0.5}
\plotone{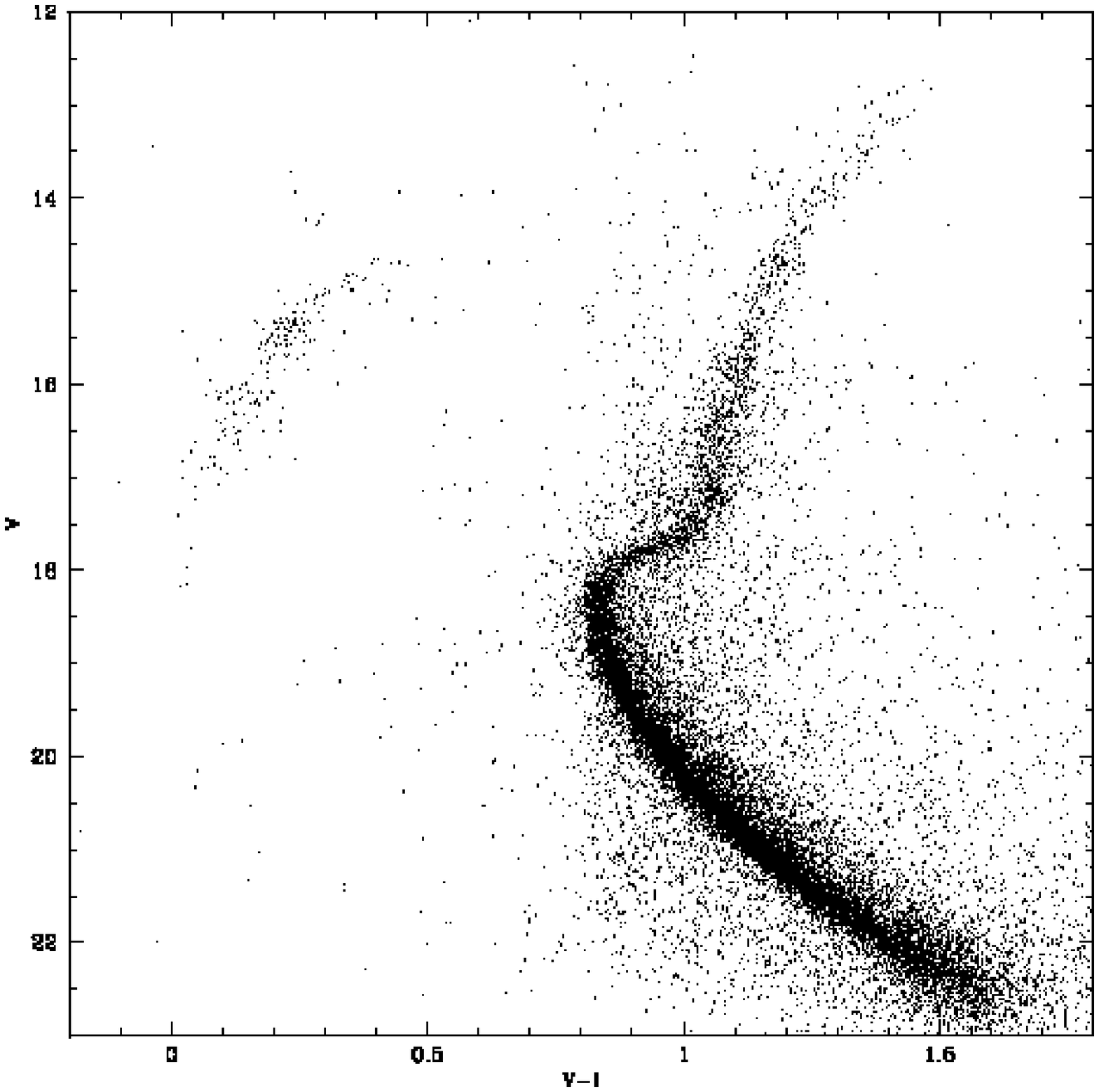} \caption{The
internally dereddened CMD of M10.  $V$ and $V-I$ indicate the
location of the data points after the differential reddening with
respect to the fiducial region was corrected for, i.e., after the
inverse values of Fig. \ref{M10_grid_lowres_trans} were applied to
the data. No reddening zero point is applied to the data in this
plot. The improvement of the appearance of the CMD over Figure
\ref{M10_rawcmd} is immediately obvious. The flaring of the data
at $V \sim 17$ is most likely due to the low signal-to-noise of
the 10s exposures at these magnitudes. \label{M10_dered_res3}}
\end{figure}

\begin{figure}
\epsscale{0.5}
\plotone{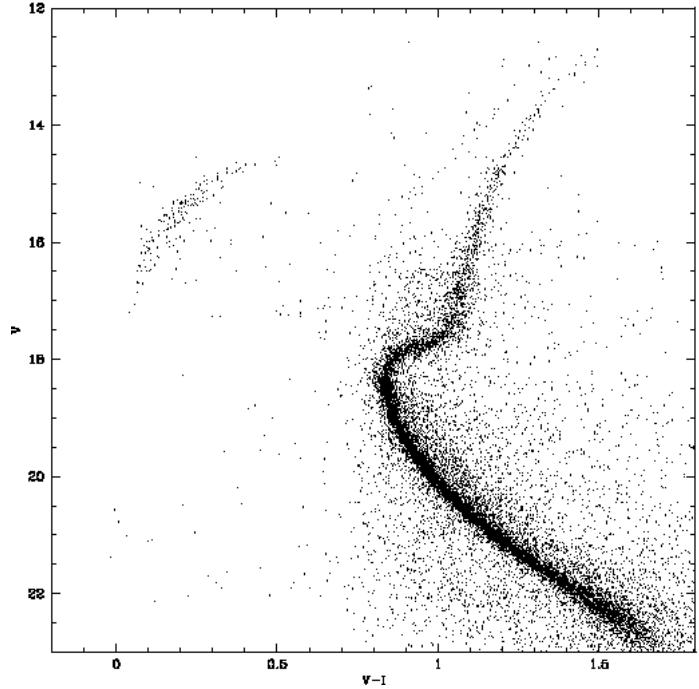} \caption{The
internally dereddened CMD of M12.  $V$ and $V-I$ indicate the
location of the data points after the differential reddening with
respect to the fiducial region was corrected for, i.e., after the
inverse values of Fig. \ref{M12_grid_lowres_trans} were applied to
the data. No reddening zero point is applied to the data in this
plot. Note the improvement of the appearance of the CMD over
Figure \ref{M12_rawcmd}. The flaring of the data at $V \sim 17$ is
most likely due to the low signal-to-noise of the 10s exposures at
these magnitudes. \label{M12_dered_res3}}
\end{figure}

The $E_{V-I}$ zero points for M10 and M12, which need to be added
to the differential $E_{V-I}$ values in the grids of Figures
\ref{M10_grid_lowres_trans} and \ref{M12_grid_lowres_trans}, are
0.23 mag and 0.24 mag, respectively, determined by fitting VDB
isochrones to the differentially dereddened data. Since previous
reddening estimates for these two GCs vary, especially for M10,
our results agree with some literature values but not with others.
Specifically, our reddening estimate for M12 agrees well with the
SFD maps, whereas our value for M10 falls below the one quoted in
SFD ($E_{V-I} \sim 0.389$), but produces by far the best VDB
isochrone fit.

Reddening studies like the one presented here may be useful in
determining properties of the interstellar medium such as a dependence
of $R_{V}$ upon position in the field of view. In addition, they may
give insight into the properties and the distribution of the dust
along the line of sight. One may argue, for instance, that M10 and
M12, since they are so close to each other in the sky, lie behind a
common, larger scale ``layer" of dust that accounts for the
practically identical reddening zero point. M10 seems to fall behind
another dust ridge \citet[suggested by, e.g.,][]{kbk96} causing the
additional differential reddening which varies on arcmin scales by
tenths of a magnitude. Furthermore, reddening maps such as the ones
presented here and in BM01 can give a sense of how small a scale
differential extinction in general may vary on, and by how much,
perhaps as a function of position in the sky or Galactic latitude.  We
hope to shed some light on this matter with a similar analysis of the
remaining GCs in our sample.

\begin{figure}
\epsscale{0.5}
\plotone{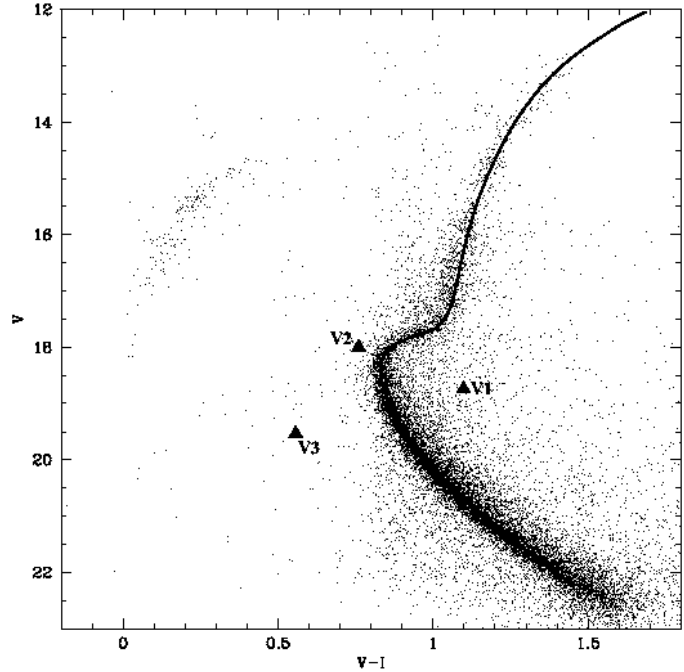}
\caption{\label{M10_isocmd} Differentially dereddened CMD of M10 with
overlain VDB best-fit isochrone and locations of variable stars. For
this fit: [Fe/H] = --1.54, age = 16 Gyrs, d = 5.1 kpc, and the
$E_{V-I}$ zero point, to which any differential extinction has to be
added, is 0.23. All CMD features are well traced out by the VDB
isochrone. The 16 Gyr isochrone produced the better fit than the 18
Gyr one, but from the appearance of the CMD, it seems that perhaps a
17 Gyr VDB isochrone, if it were available, would produce an even
better fit than the one shown here.  We also show in this Figure the
locations of the variable star candidates in the CMD of M10. The data
presented, including the variable stars, are dereddened to the M10
fiducial region. No reddening zero point is applied. The variables are
plotted at maximum brightness. From this figure alone, it seems that
none of the variables we found in the field of M10 are members of the
cluster. It should be noted that since we only use our 600s data for
binary detection, we are not sensitive to variables brighter than $V
\sim 16.5$.}
\end{figure}

\begin{figure}
\epsscale{0.5}
\plotone{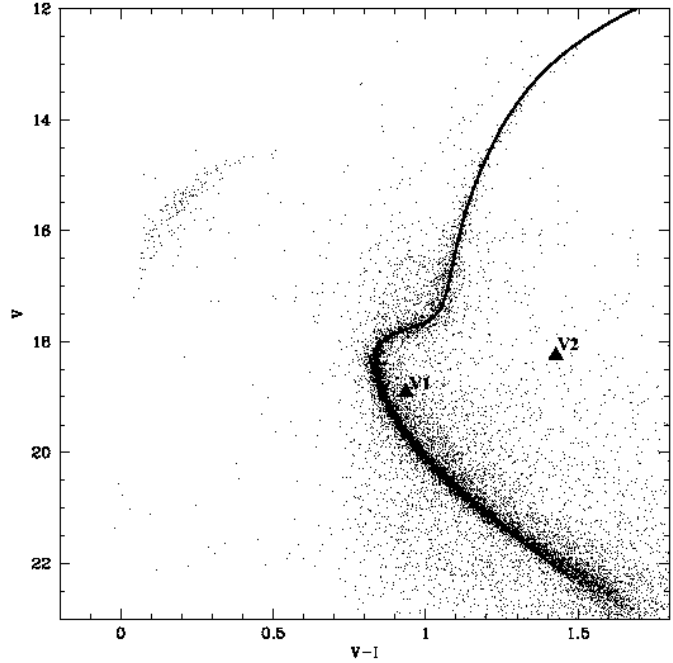}
\caption{\label{M12_isocmd} Differentially dereddened CMD of M12 with
overlain VDB best-fit isochrone and locations of variable stars. For
this fit: [Fe/H] = --1.54, age = 16 Gyrs, d = 4.9 kpc, and the
$E_{V-I}$ zero point is 0.24. All features of the CMD, including the
subgiant branch and the RGB, are very well traced out. The only minor
deviation between the loci of the data points and the isochrone fit
occurs at $V \sim 21.5$. The above isochrone parameters agree with the
values for M12 in \citet{harris1996} and the SFD $E_{V-I}$ estimates
for this region.  Also shown are the locations of the variable star
candidates in the CMD of M12. The data presented, including the
variable stars, are dereddened to the M12 fiducial region. No
reddening zero point is applied. The variables are plotted at maximum
brightness. From this figure alone, it seems that the variable V1 is a
cluster member whereas V2 is not. It should be noted that since we
only use our 600s data for binary detection, we are not sensitive to
variables brighter than $V \sim 16.5$.}
\end{figure}

\begin{figure}
\epsscale{0.5}
\plotone{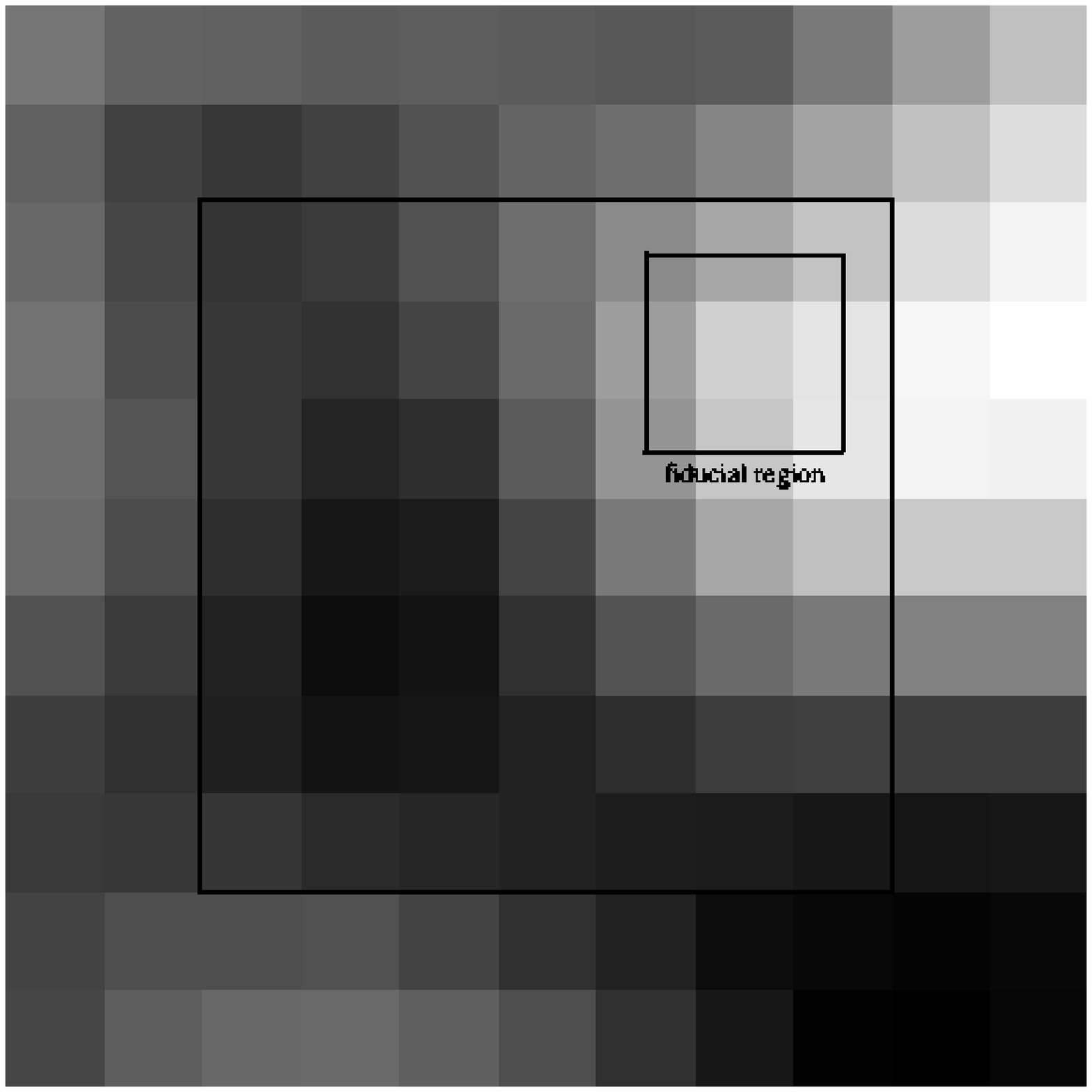}
\caption{\label{M10_sfd_lowres}Graphical Representation of the SFD
Map in the region of M10. The location of the fiducial region is
given for reference. Orientation, size, and pixel sizes are the
same as in Figures \ref{M10_extmap_lowres} and
\ref{M10_grid_lowres_trans}. Darker regions correspond to higher
extinction. The extinction ranges from $E_{V-I} = 344$ to 411
mmag. The average reddening is $E_{V-I} = 387 \pm 17$ mmag for the
entire field and $E_{V-I} = 389 \pm 16$ mmag for the inner part of
the field. }
\end{figure}

\begin{figure}
\epsscale{0.5}
\plotone{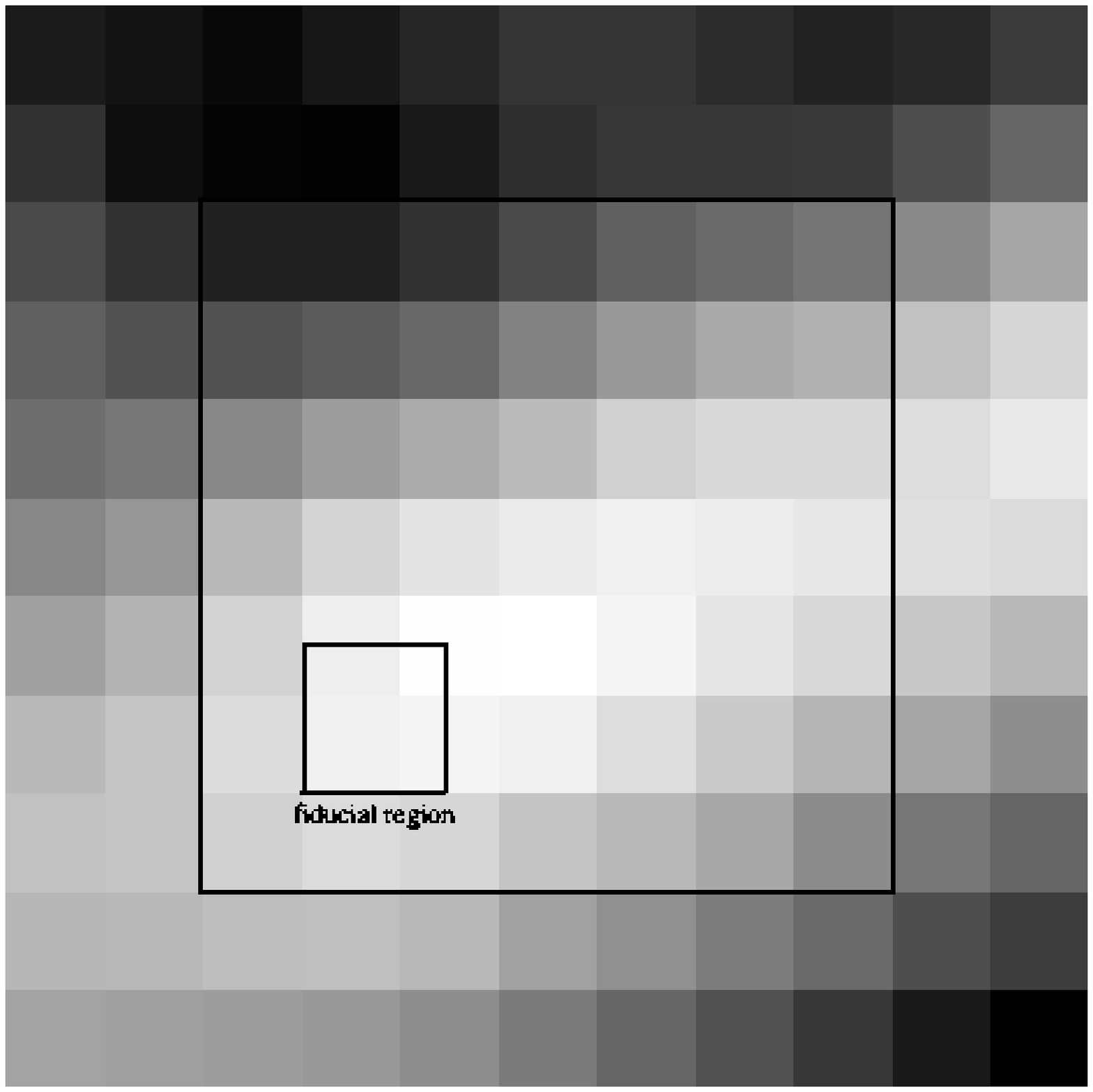}
\caption{\label{M12_sfd_lowres}Graphical Representation of the SFD
Map in the region of M12. The location of the fiducial region is
given for reference. Orientation, size, and pixel sizes are the
same as in Figures \ref{M12_extmap_lowres} and
\ref{M12_grid_lowres_trans}. Darker regions correspond to higher
extinction. The extinction ranges from $E_{V-I} = 241$ to 271
mmag. The average reddening is $E_{V-I} = 254 \pm 8$ mmag for the
entire field and $E_{V-I} = 250 \pm 7$ mmag for the inner part of
the field. }
\end{figure}

\begin{figure}
\epsscale{0.5}
\plotone{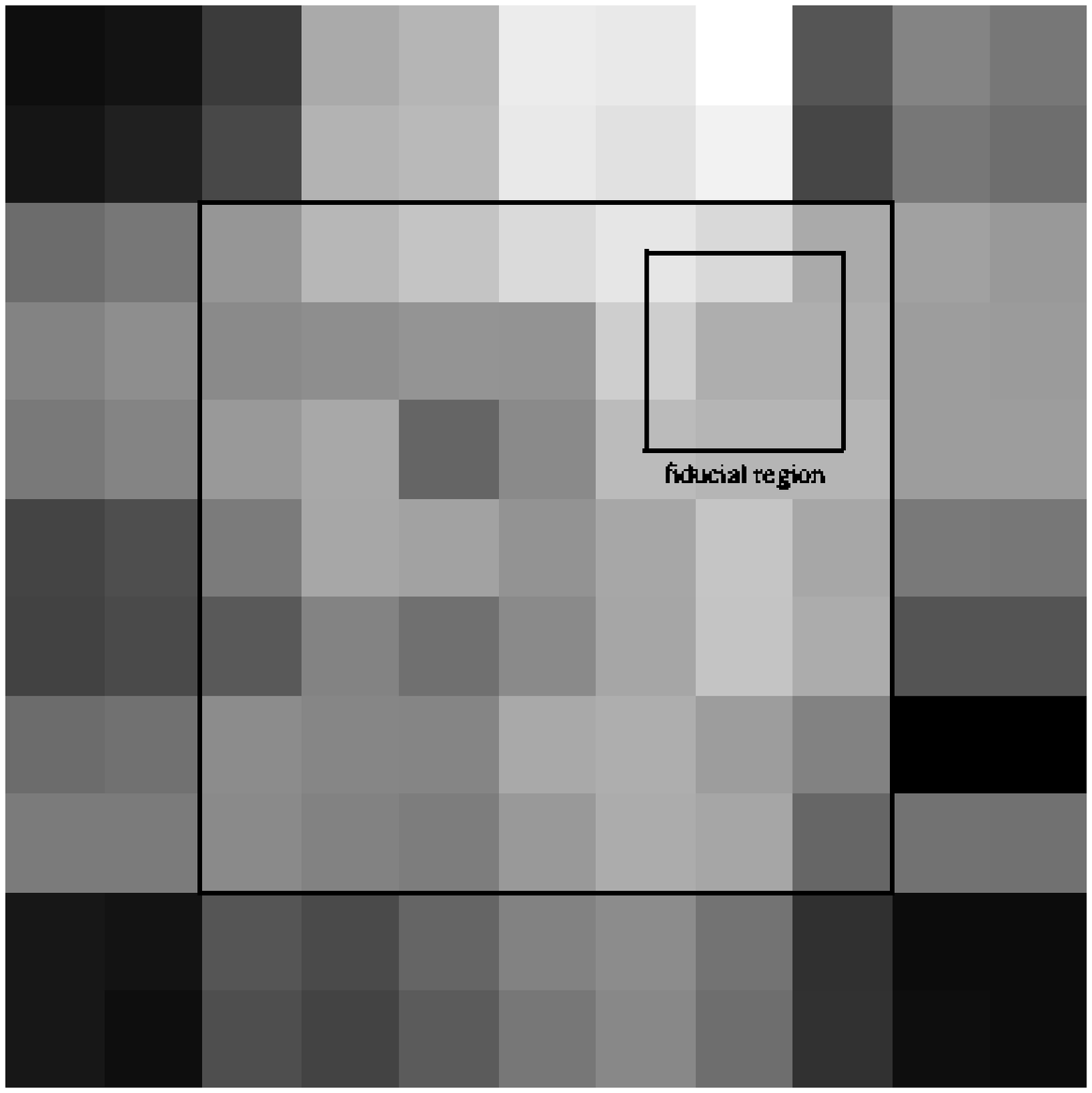}
\caption{\label{M10_diffmap_lowres}Graphical Representation of the
difference between our extinction map for M10 (Fig.
\ref{M10_extmap_lowres}) and the SFD Map of the same region (Fig.
\ref{M10_sfd_lowres}). The location of the fiducial region is
given for reference; the orientation, size, and pixel sizes are
the same as in Figures \ref{M10_extmap_lowres} and
\ref{M10_grid_lowres_trans}. Darker regions correspond to areas
where our maps indicate more differential extinction (relative to
the fiducial region) than the SFD map. It is obvious that the
especially the inner region is featureless, indicating good
spatial agreement between the two maps. The pixels toward the
edges of the map contain fewer cluster stars causing any
calculated differential reddening to have a higher associated rms
error which results in the ``noisy" appearance of the ring around
the inner region. The average reddening of the whole field is
$E_{V-I} = 319 \pm 35$ mmag and $E_{V-I} = 340 \pm 16$ mmag for
the inner region.}
\end{figure}

\begin{figure}
\epsscale{0.5}
\plotone{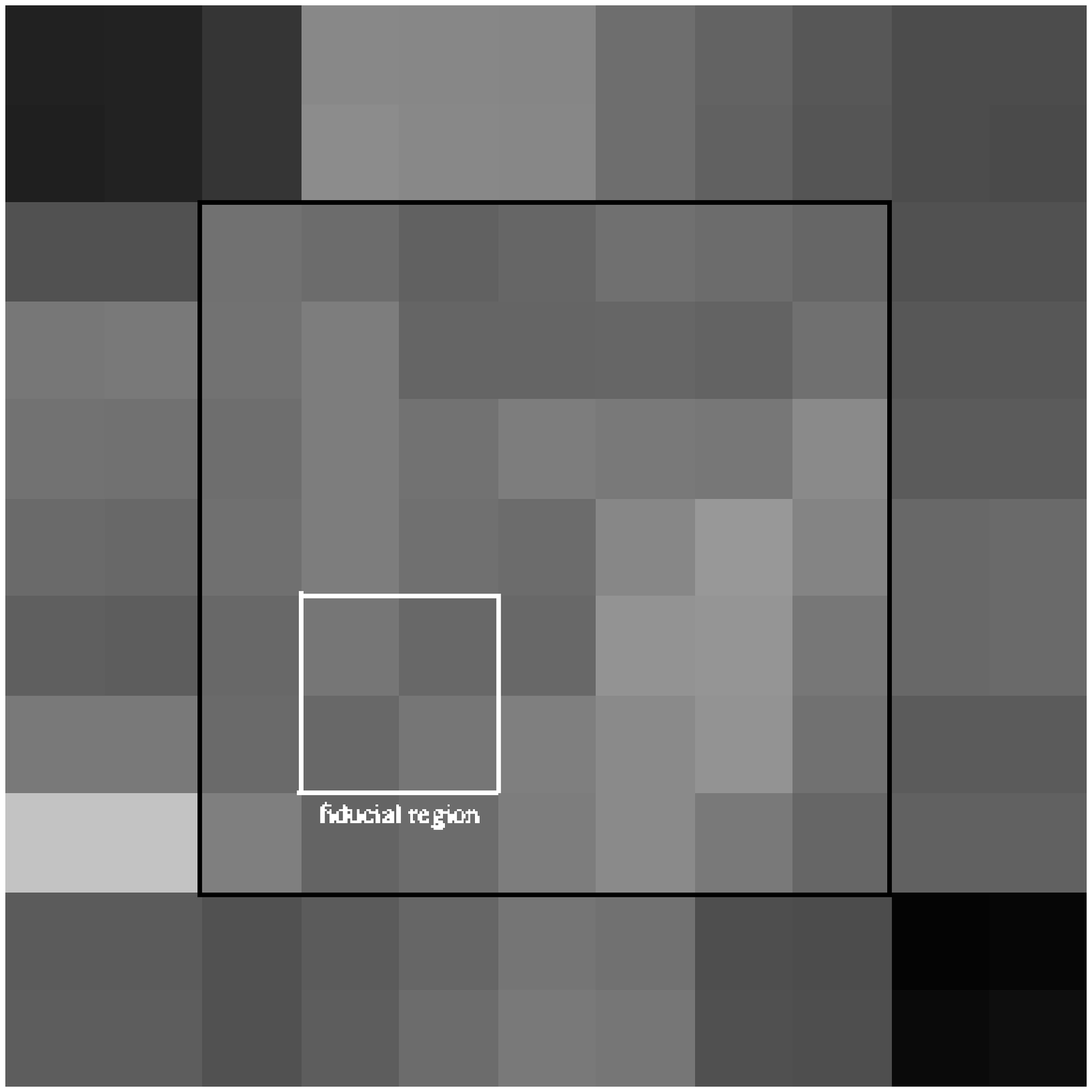}
\caption{\label{M12_diffmap_lowres}Graphical Representation of the
difference between our extinction map for M12 (Fig.
\ref{M12_extmap_lowres}) and the SFD Map of the same region (Fig.
\ref{M12_sfd_lowres} . The location of the fiducial region is
given for reference, the orientation, size, and pixel sizes are
the same as in Figures \ref{M12_extmap_lowres} and
\ref{M12_grid_lowres_trans}. Darker regions correspond to areas
where our map indicates more differential extinction (relative to
the fiducial region) than the SFD map. The difference map is
featureless except in the regions towards the corner of the image
which is evidence for the good spatial agreement between the two
maps. The average reddening of the whole field is $E_{V-I} = 214
\pm 53$ mmag, whereas for only the inner region, it is $E_{V-I} =
241 \pm 11$ mmag.}
\end{figure}

\begin{figure}
\epsscale{0.5}
\plotone{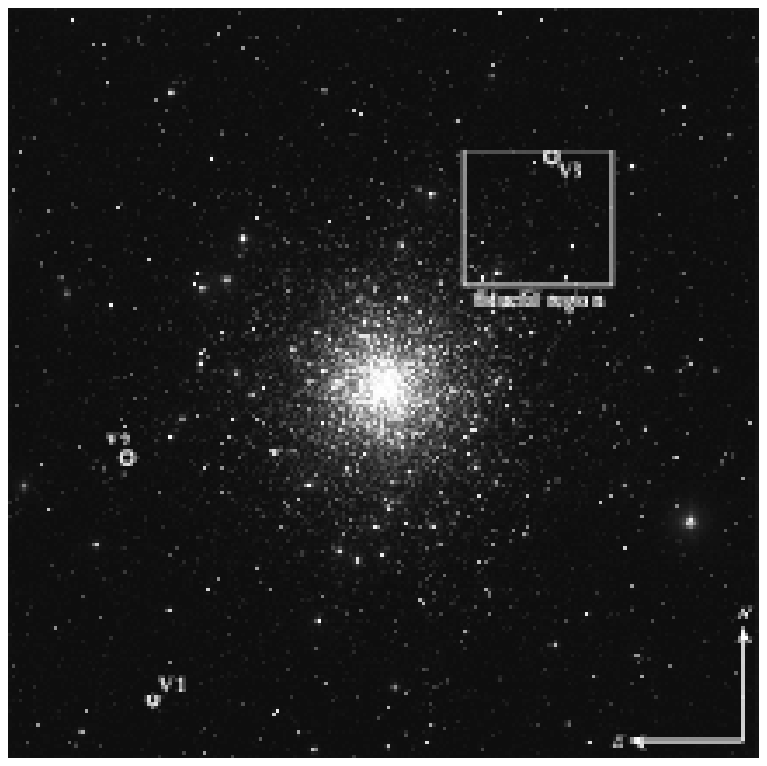}
\caption{\label{M10_field}CCD Field of view of M10 with the
locations of the variables. The fiducial region is shown for
reference. The size of the field of view is 81\% (23.5 arcmin on
the side) of the area shown in Figures \ref{M10_extmap_lowres},
\ref{M10_grid_lowres_trans}, \ref{M10_sfd_lowres}, and
\ref{M10_diffmap_lowres}.}
\end{figure}

\begin{figure}
\epsscale{0.5}
\plotone{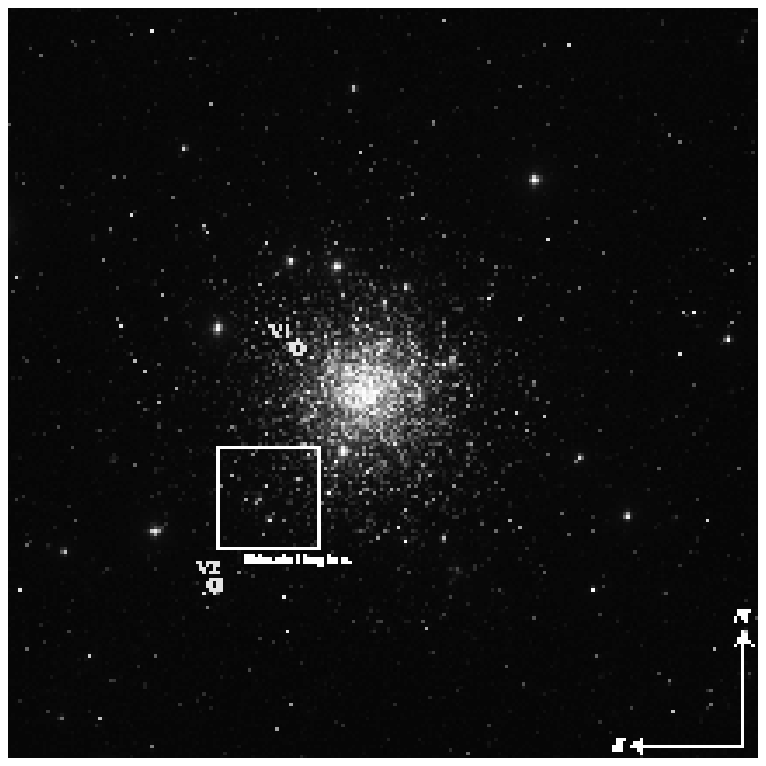}
\caption{\label{M12_field}CCD Field of view of M12 with the
locations of the variables. The fiducial region is shown for
reference. The size of the field of view is 81\% (23.5 arcmin on
the side) of the area shown in Figures \ref{M12_extmap_lowres},
\ref{M12_grid_lowres_trans}, \ref{M12_sfd_lowres}, and
\ref{M12_diffmap_lowres}.}
\end{figure}

\begin{figure}
\epsscale{0.5}
\plotone{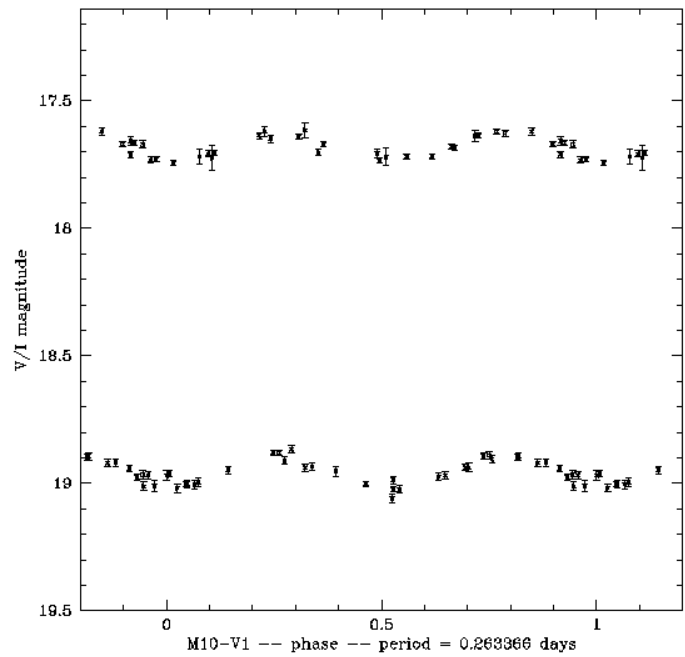}
\caption{\label{M10_v1}V1 in the field of M10, a W UMa binary system
with a period of 0.263366 days. $I$ data are plotted above the $V$
data. No reddening correction is applied to the lightcurve
data. M10-V1 is not a member of M10.}
\end{figure}

\begin{figure}
\epsscale{0.5}
\plotone{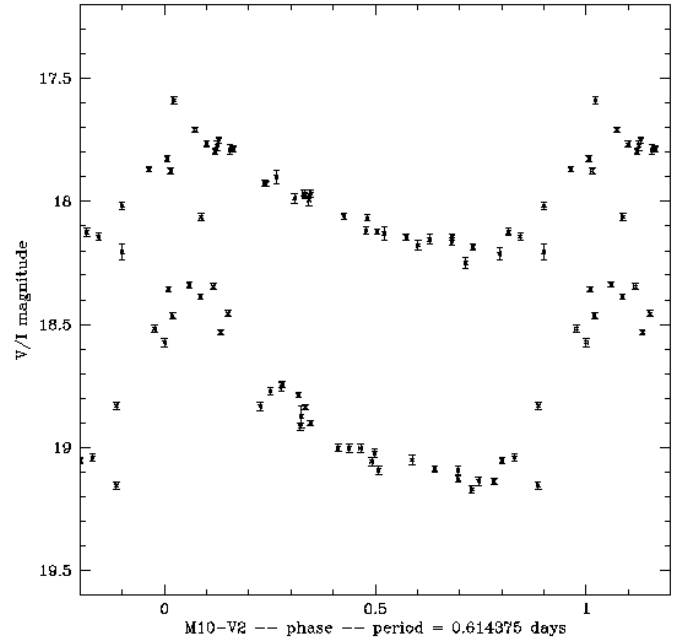}
\caption{\label{M10_v2}V2 in the field of M10, an RR Lyrae system with
a period of 0.61438 days. $I$ data are on top.  No reddening
correction is applied. The variable is a background star (40 kpc).}
\end{figure}

\begin{figure}
\epsscale{0.5}
\plotone{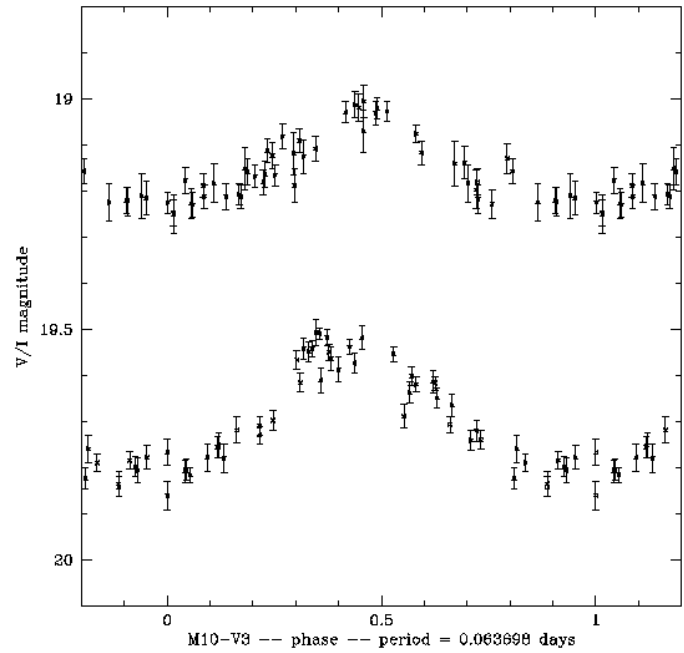}
\caption{\label{M10_v3}V3 in the field of M10, an SX Phe pulsating
variable with a period of 0.0637 days.  $V$ data are plotted below the
$I$ data. No reddening correction is applied to the lightcurve
data. M10-V3 is a background field star (distance $\sim$ 16 kpc).}
\end{figure}

\begin{figure}
\epsscale{0.5}
\plotone{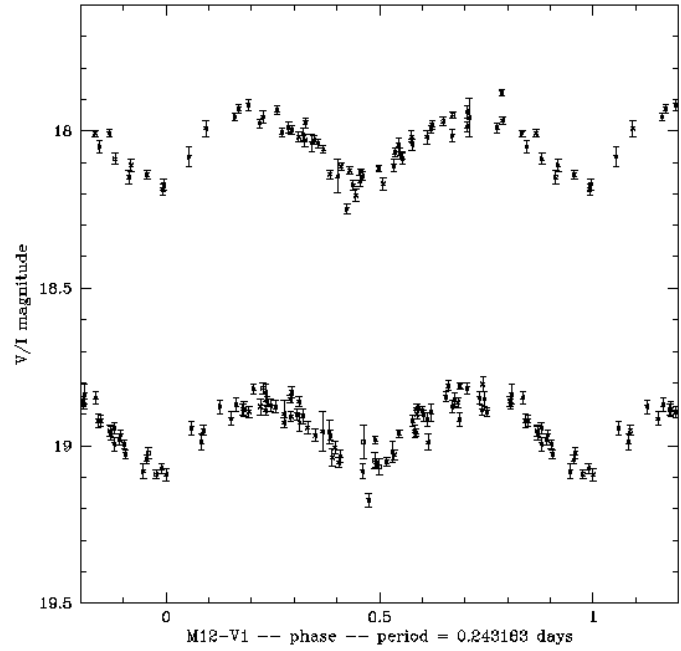}
\caption{\label{M12_v1}V1 in the field of M12, a W UMa binary system
with a period of 0.243183 days and the only variable which is most
likely physically associated with a GC. $I$ data are above the $V$
data. No reddening correction is applied to the lightcurve data.}
\end{figure}

\begin{figure}
\epsscale{0.5}
\plotone{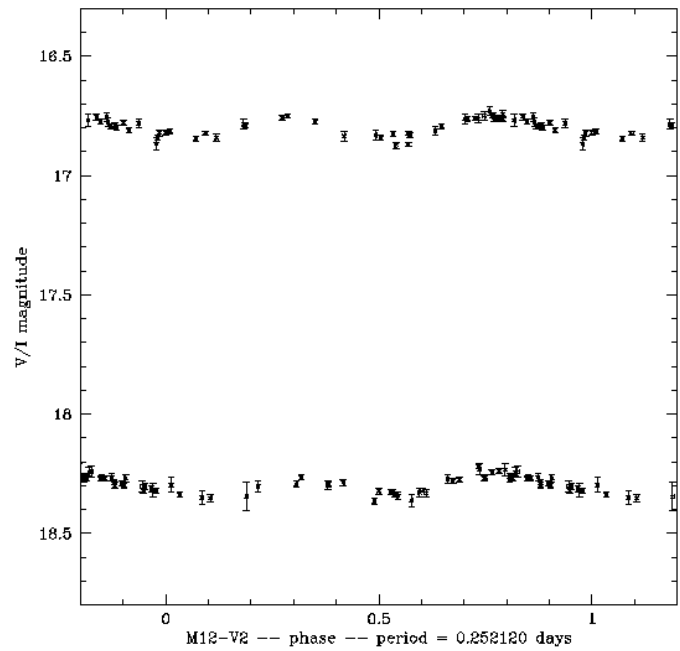}
\caption{\label{M12_v2}V2 in the field of M12, a W UMa binary system
with a period of 0.25212 days. $I$ data are plotted above the $V$
data. No reddening correction is applied to the lightcurve data. }
\end{figure}

Our search for variable stars in M10 and M12 produced five
candidates. In M10, we detected a W UMa binary, an RR Lyrae, and an SX
Phe field variable. None of these three systems appears to be
physically associated with M10, given their locations in the CMD
(Figure \ref{M10_isocmd}) and their estimated distances (Table 7 and
Section 4.3.3). In M12, we detected two W UMa binary systems, one of
which (M12-V1) is most likely a cluster member.  Again, we base this
estimate on CMD location (Figure \ref{M12_isocmd}) and calculated
distance moduli (Table 7).

As in the case of NGC 3201 (BM02), we detect a number of foreground
and background variables but only one (likely) GC member binary. Both
M10 and M12 are at higher Galactic latitude ($\sim 25^{\circ}$) than
NGC 3201 ($\sim 8.5^{\circ}$), so the disk contamination is not as
significant. The number of total stars which were monitored was only
slightly different (27000 for NGC 3201 and 34000 for M10 and M12
combined), but in the case of NGC 3201, we found 14 variables (11
binaries, 1 member), whereas in this work we find 5 variables (3
binaries, 1 probable member).  In both studies, cluster-member stars
outnumber non-member stars, as may easily be seen in the CMDs of the
clusters. Yet the number of non-member binaries we find is much higher
than the number of member binaries\footnote{It should be noted that we
are not sensitive to cluster-member pulsating variables since almost
all of them are saturated in the data we analyze for the existence of
binaries. A comparison between GC member and non-member pulsating
variables is there not meaningful.}. It will be interesting to see
whether this trend continues in the rest of the GCs in our sample
which will be analyzed using the same methods as outlined in this
work, BM01, and BM02, or whether NGC 3201, M10, and M12 are merely
``special cases".

Finally, we have shown that we are able to detect binaries of very low
amplitude ($\sim 0.07$ mag; BM02) and duty cycle (0.1; BM02) and of
short ($\sim 0.13$ days\footnote{M10-V3 actually has a period of just
above 0.06 days, but only has one maximum per period. Were it a binary
star, it would have two maxima, and its period would consequently be
around 0.13 days}; this work) and long periods (2.85 days; BM02), even
when using very different observing techniques for the clusters. Using
our existing 1m-class telescope photometry data set of the rest of the
GCs in our sample, we should be able to identify binaries and other
variables in these clusters (provided they exist) and determine their
periods. With the spectroscopic capabilities of modern 8m-class
telescopes for follow-up observations, we can determine cluster
membership for these variables in the near future.

%--------------------------------------------------------------------------------

%--------------------------------------------------------------------------------
\acknowledgements{}

This research was funded in part by NSF grants AST 96-19632 and
98-20608. Thanks to the anonymous referee for insightful comments and
suggestions. We would also like to thank Louis Strolger for his help
with the world coordinate system Matching. Furthermore, we thank
R. Dohm-Palmer for providing us with algorithms to isolate stars in
the fiducial regions of our extinction maps based on their CMD
locations. Special thanks also to Ian Thompson for his help with the
LCO filter transmission curves, to Doug Finkbeiner for pointing out
several pitfalls in the process of the extinction calculation, and to
Andrzej Udalski for help with the AoV algorithm.  Finally, we would
like to express our most sincere gratitude to the support staff at MDM
and LCO for their countless hours of help and assistance.
%--------------------------------------------------------------------------------

%--------------------------------------------------------------------------------

%--------------------------------------------------------------------------------

%--------------------------------------------------------------------------------
% Tables
%--------------------------------------------------------------------------------
\clearpage

% Table 1 - deluxetable format - enclosed

\begin{deluxetable}{cccc}
\tabletypesize{\scriptsize} \tablecaption{M10 Observations}
\tablewidth{0pt} \tablehead{ \colhead {Date} & \colhead
{Telescope} & \colhead {Field-of-View Size\tablenotemark{a}} &
\colhead {Number of Epochs\tablenotemark{b}}} \startdata
April 1996 & MDM 1.3m & 14.9 & 4\\
May 1997 & LCO 1m & 23.5 & 29\\
June 1997 & MDM 1.3m & 14.9 & 35\\
August 1997 & LCO 1m & 23.5 & 23\\
May 1998\tablenotemark{c} & LCO 1m & 23.5 & 25\\
August 1998 & MDM 1.3m & 17 & 16\\
August 1998 & LCO 1m & 23.5 & 14\\
May 1999 & MDM 1.3m & 17 & 12\\
\enddata

\tablenotetext{a}{In units of arcminutes on the side.}
\tablenotetext{b}{This number of 600s exposure time epochs is more
or less evenly divided between $V$ and $I$.}
\tablenotetext{c}{During this run, we took additional 10s
exposures to complete the CMD in brighter regions. These shorter
exposures were not used to look for variables.}

\end{deluxetable}
%--------------------------------------------------------------------------------
%\clearpage

% Table 2 - deluxetable format - enclosed

\begin{deluxetable}{cccc}
\tabletypesize{\scriptsize} \tablecaption{M12 Observations}
\tablewidth{0pt} \tablehead{ \colhead {Date} & \colhead
{Telescope} & \colhead {Field-of-View Size\tablenotemark{a}} &
\colhead {Number of Epochs\tablenotemark{b}}} \startdata
June 1995 & MDM 1.3m & 10.6 & 60\\
April 1996 & MDM 1.3m & 14.9 & 4\\
May 1997 & LCO 1m & 23.5 & 28\\
June 1997 & MDM 1.3m & 14.9 & 39\\
August 1997\tablenotemark{c} & LCO 1m & 23.5 & 18\\
May 1998 & LCO 1m & 23.5 & 22\\
August 1998 & MDM 1.3m & 17 & 16\\
August 1998 & LCO 1m & 23.5 & 8\\
May 1999 & MDM 1.3m & 17 & 10\\
\enddata

\tablenotetext{a}{In units of arcminutes on the side.}
\tablenotetext{b}{This number of 600s exposure time epochs is more
or less evenly divided between $V$ and $I$.}
\tablenotetext{c}{During this run, we took additional 10s
exposures to complete the CMD in brighter regions. These shorter
exposures were not used to look for variables.}

\end{deluxetable}
%--------------------------------------------------------------------------------
%\clearpage

% Table 3 - deluxetable format - enclosed

\begin{deluxetable}{ccc}
\tabletypesize{\scriptsize} \tablecaption{Selected Recent Average
$E_{V-I}$ Estimates for M10 } \tablewidth{0pt} \tablehead{
\colhead {$E_{V-I}$} & \colhead {Reference} & \colhead {Method}}

\startdata

214 mmag & \citet{bh82} & HI/galaxy count maps\\
426 mmag & \citet{rhs88} & integrated light\\
371 mmag & \citet{hrf89} & $UBV$ 2-color diagram\\
234 mmag & \citet{amp90} & infrared photometry\\
382 mmag & \citet{harris1996} & compilation of selected previous
results\\
389 mmag\tablenotemark{a} & \citet{sfd98} & dust IR emissivity maps\\
279 mmag & this work & VDB isochrone fitting\\

\enddata

\tablenotetext{a}{This value also corresponds to our estimate
using the SFD maps.} \tablecomments{All listed literature
reddening estimates were obtained by converting $E_{B-V}$ values
to $E_{V-I}$ by using $E_{B-V} = 0.7273 \times E_{V-I}$ and a
$R_{V} = 3.1$ reddening law \citep{sfd98}.}

\end{deluxetable}
%--------------------------------------------------------------------------------
%\clearpage

% Table 4 - deluxetable format - enclosed

\begin{deluxetable}{ccc}
\tabletypesize{\scriptsize} \tablecaption{Selected Recent Average
$E_{V-I}$ Estimates for M12} \tablewidth{0pt} \tablehead{ \colhead
{$E_{V-I}$} & \colhead {Reference} & \colhead {Method}}

\startdata

127 mmag & \citet{bh82} & HI/galaxy count maps\\
317 mmag & \citet{srf89} & $UBV$ 2-color diagram\\
258 mmag & \citet{harris1996} & compilation of selected previous
results\\
250 mmag\tablenotemark{a} & \citet{sfd98} & dust IR emissivity maps\\
249 mmag & this work & VDB isochrone fitting\\

\enddata

\tablenotetext{a}{This value also corresponds to our estimate
using the SFD maps.} \tablecomments{All listed literature
reddening estimates were obtained by converting $E_{B-V}$ values
to $E_{V-I}$ by using $E_{B-V} = 0.7273 \times E_{V-I}$ and a
$R_{V} = 3.1$ reddening law \citep{sfd98}.}

\end{deluxetable}
%--------------------------------------------------------------------------------
%\clearpage

% Table 5 - deluxetable format - enclosed

\begin{deluxetable}{ccc}
\tabletypesize{\scriptsize} \tablecaption{Selected Recent Average
$E_{V-I}$ Estimates for NGC 3201} \tablewidth{0pt} \tablehead{
\colhead {$E_{V-I}$} & \colhead {Reference} & \colhead {Method}}

\startdata

285 mmag & \citet{c84} & RR Lyrae colors\\
285 mmag & \citet{harris1996} & compilation of selected previous
results\\
336 mmag\tablenotemark{a} & \citet{sfd98} & dust IR emissivity maps\\
240 mmag & \citet{BM01} & VDB isochrone fitting\\

\enddata

\tablenotetext{a}{This value also corresponds to our estimate
using the SFD maps.} \tablecomments{All listed literature
reddening estimates were obtained by converting $E_{B-V}$ values
to $E_{V-I}$ by using $E_{B-V} = 0.7273 \times E_{V-I}$ and a
$R_{V} = 3.1$ reddening law \citep[$\lambda_{central} \sim 8100$
\AA]{sfd98}. They represent the updated values from BM01 where we
calculated $E_{B-V} = 0.6468 \times E_{V-I}$, based on
\citet[$\lambda_{central} \sim 9000$ \AA]{ccm89}.}

\end{deluxetable}
%--------------------------------------------------------------------------------
%\clearpage

% Table 6 - deluxetable format - enclosed

\begin{deluxetable}{cccclcc}
\tabletypesize{\scriptsize} \tablecaption{Variable Stars in the
Fields of M10 \& M12} \tablewidth{0pt} \tablehead{ \colhead{Var.
No.} & \colhead{type} & \colhead{RA (2000)} & \colhead{Dec (2000)}
& \colhead{period (days)} & \colhead{$V_{bright}$} &
\colhead{$I_{bright}$} } \startdata M10-V1 & W UMa & 16:57:39.15 &
--4:16:03.9& 0.263366(28)& 18.890(17)& 17.620(19)\\
M10-V2& RR Lyrae & 16:57:41.57 & --4:08:24.8 & 0.61438(23) &
18.170(14)
& 17.650(19)\\
M10-V3& SX Phe & 16:56:48.63 & --3:58:35.9 & 0.0637(41){*} &
19.538(18)& 19.013(33)\\
M12-V1 & W UMa& 16:47:22.89 & --1:55:35.8 & 0.243183(15) &
18.818(19) & 17.949(19)\\
M12-V2 & W UMa& 16:47:32.58 & --2:03:10.1 & 0.25212(25){*} &
18.238(12) & 16.750(19)\\
\enddata

\tablecomments{
\begin{itemize}
\item Errors in parentheses indicate the uncertainty in last two digits.
\item Photometry errors are the result of adding in quadrature the DoPHOT
photometry error for the instrumental magnitude and the rms of the
standard star solution (see BM01).
\item The random error in the period corresponds to the full width at half maximum
(fwhm) of the peak in the AoV power spectrum corresponding to the
correct frequency. For the determination of this error, only $V$
data were used. For the variables M10-V3 and M12-V2, the peak in
the power spectrum corresponding to the correct period was
assigned essentially the same power as the directly neighboring
peak (i.e., it would be hard to pick ``by eye'' which one is the
correct one). In these instances, we estimated the period rms to
be the distance between these two neighboring peaks. The two cases
are marked by an asterisk.
\end{itemize}
}

\end{deluxetable}
%--------------------------------------------------------------------------------
%\clearpage

% Table 7 - deluxetable format - enclosed

\begin{deluxetable}{clcccc}
\tabletypesize{\scriptsize} \tablecaption{Rucinski Magnitudes and
Distance Moduli for W UMa-type Binaries in the Fields of M10 \&
M12} \tablewidth{0pt} \tablehead{ \colhead {system} & \colhead
{period (days)\tablenotemark{a}} & \colhead
{$E_{V-I}$\tablenotemark{b}} & \colhead {$(V-I)_{0,
bright}$\tablenotemark{c}} & \colhead
{$M_{V(Rucinski)}$\tablenotemark{d}} & \colhead
{$V_{0}-M_{V}$\tablenotemark{e}} }

\startdata

M10-V1 & 0.263366(28) & 0.323(07) & 0.947 & 5.695&
12.576\\
M12-V1 & 0.243183(15) & 0.249(04) & 0.620 & 4.661&
13.679\\
M12-V2 & 0.25212(25) & 0.259(03) & 1.229 & 6.802&
10.939\\

\enddata

\tablenotetext{a}{As in Table 3, errors in parentheses indicate
the uncertainty in last two digits. The period errors are the same
as in Table 3.} \tablenotetext{b}{$E_{V-I}$ contains the reddening
zero points for the two clusters, calculated in Section 3.2.1. As
mentioned above, this value assumes that the binary suffers the
full extinction along the line of sight to the GC (which is not
necessarily correct if the binary is not a cluster member).  The
errors for $E_{V-I}$ are the random errors in the determination of
the \textit{differential reddening.} That is, any possible
systematic error in the determination of the reddening zero point
is not included in this estimate.} \tablenotetext{c}{$(V-I)_{0,
bright}$ is the dereddened color at maximum light.}
\tablenotetext{d}{$M_{V}=-4.43\log P+3.63 (V-I)_{0}-0.31$ (see
Section 4.3.3)}. \tablenotetext{e}{The true distance modulus to
M10 is $V_{0}-M_{V}=13.55$ and the one to M12 is 13.46. These
values represent the apparent distance moduli corrected for
extinction.}

\tablecomments{ Rucinski quotes the scatter in his relation
\citep{rucinski1994,rucinski1995} to be 0.29 mag in the
calculation of the absolute $V$ magnitudes (corresponding to an
uncertainty of approximately 13\% in distance). Since this
uncertainty is far larger than the quadratic sum of all our random
errors, we refrain from a detailed error analysis for the Rucinski
magnitudes and distance moduli.  }

\end{deluxetable}
%--------------------------------------------------------------------------------

%--------------------------------------------------------------------------------
% Figures
%--------------------------------------------------------------------------------
%\clearpage

%--------------------------------------------------------------------------------
\clearpage

%--------------------------------------------------------------------------------
%\clearpage

%--------------------------------------------------------------------------------
%\clearpage

%--------------------------------------------------------------------------------
%\clearpage

%--------------------------------------------------------------------------------
%\clearpage

%--------------------------------------------------------------------------------
%\clearpage

%--------------------------------------------------------------------------------
%\clearpage

%--------------------------------------------------------------------------------
%\clearpage

%--------------------------------------------------------------------------------
%\clearpage

%--------------------------------------------------------------------------------
%\clearpage

%--------------------------------------------------------------------------------
%\clearpage

%--------------------------------------------------------------------------------
%\clearpage

%--------------------------------------------------------------------------------
%\clearpage

%--------------------------------------------------------------------------------
%\clearpage

%--------------------------------------------------------------------------------
%\clearpage

%--------------------------------------------------------------------------------
%\clearpage

%--------------------------------------------------------------------------------
\clearpage

%--------------------------------------------------------------------------------
%\clearpage

%--------------------------------------------------------------------------------
%\clearpage

%--------------------------------------------------------------------------------
%\clearpage

%--------------------------------------------------------------------------------
\end{document}